\input phyzzx

\def\stackrel#1#2{\mathrel{\mathop{#2}\limits^{#1}}}

\def\Haq{{\widetilde{\cal H}}^{\hbar}}
\Pubnum={$\rm UTS-DFT-93-10$}
\date={}
\pubtype={}
\titlepage
\title{A PROPOSAL FOR A DIFFERENTIAL CALCULUS IN QUANTUM MECHANICS}
\author{E.Gozzi$^{\flat}$ and M.Reuter$^{\sharp}$}
\address{$\flat$ Dipartimento di Fisica Teorica, Universit\`a di Trieste,\break
Strada Costiera 11, P.O.Box 586, Trieste, Italy \break and INFN, Sezione 
di Trieste.\break
\break
$\sharp$ Deutsches Elektronen-Synchrotron DESY,\break Notkestrasse 85, 
W-2000 Hamburg 52, Germany}
\abstract
In this paper, using the Weyl-Wigner-Moyal formalism for quantum mechanics,
we develop a {\it quantum-deformed} exterior calculus on the phase-space of an
arbitrary hamiltonian system. Introducing additional bosonic and fermionic
coordinates we construct a super-manifold which is closely related
to the tangent and cotangent bundle over  phase-space.
Scalar functions on the super-manifold become equivalent to differential forms
on the standard phase-space. The algebra of these functions is equipped with 
a Moyal super-star product which deforms the pointwise product of the
classical tensor calculus. We use the Moyal bracket algebra in order to
derive a set of quantum-deformed rules for the exterior derivative, 
Lie derivative, contraction, and similar operations of the Cartan
calculus.

\endpage
\chapter{INTRODUCTION}
The physics community has recently witnessed a growing interest in quantum groups
\Ref\polac{V.Drinfeld, in:{\it "Proceedings of the International Congress of 
Mathematicians"} (Berkeley), Acad.Press.,~Vol. 1 (1986) 798;\nextline
M.Jimbo, Lett.Math.Phys. 11 (1986) 247}. These are deformations of classical
Lie algebras which first appeared in the context of the quantum inverse 
scattering method\Ref\fad{L.D.Faddeev, N.Y.Reshetikhin, L.A.Takhtajian, Preprint
LOMI E-14-87}. A lot of efforts\Ref\Woro{S.L.Woronowicz, Comm.Math.Phys.
122 (1989) 125\nextline
D.Bernard, {"\it Quantum Lie Algegras and differential calculus on quantum
groups"}\nextline
Saclay-spht-90-124, Sept 1990.\nextline
To appear in the proceedings of the 1990 Yukawa international seminar
"{\it Common trends in mathematics and quantum field theory}",
Kyoto 1990} has already gone into the investigation
of the differential structures associated to quantum groups, but the program
of developing a quantum deformed differential calculus  and investigating 
its impact on physics is certainly still in its infancy. It became clear by
now that there is a close relationship between quantum groups and the general
framework of non-commutative geometry\Ref\conn{A.Connes,~"{\it Non-commutative
differential geometry}",~Publ.Math.IHES, 62 (1985) 41}\Ref\man{Yu.I.Manin,
"{\it Quantum groups and non-commutative geometry}", \nextline
Montreal Univ. CRM-1561 (1988)}~which, loosely speaking, deals with spaces whose coordinates are
non-commuting objects. It is one of the basic credos of non-commutative
geometry that these spaces should not be investigated by visualizing them as
a set of points, but rather by studying the algebra of functions defined on
them. An important example \Ref\usd{J.Wess and B.Zumino, Nucl.Phys. B 
(Proc.Suppl.) 18 (1990) 302;
\nextline
M.Dubois-Violette, in "{\it Differential Geometric Methods in Theoretical
Physics"}, \nextline C.Bertocci et al.~(Eds.),~Lecture-Notes in Physics,
 no.275, Springer, New York (1991);
\nextline
A.Dimakis and F.M\"uller-Hoissen, J.Phys.A 25 (1992) 5625}
of a non-commutative manifold which can be 
investigated in this framework is the quantum mechanical phase-space.~
Canonical quantization turns the c-number coordinates of the classical
phase-space into non-commuting operators so that it is not clear a priori
in which sense quantum  phase-space can be considered a "manifold".
A first step towards a non-commutative geometry of phase-space was 
taken long ago by Moyal\Ref\moyo{J.E.Moyal, Proc.Cambridge Phil.Soc.
45 (1949) 99} and by Bayen et al.\Ref\baye{F.Bayen et al.,~Ann.of Phys.
111 (1978) 61; ibid. 111 (1978) 111} who, building upon the work
of Weyl and Wigner\Ref\wig{H.Weyl, Z.Phys.46 (1927) 1;\nextline
E.Wigner, Phys.Rev.40 (1932) 740}, reformulated quantum mechanics in
terms of functions on phase-space. The concept of "quantum" phase-space
employed here is classical in the sense that the coordinates commute, but 
a non-classical feature is introduced via a new non-commutative
product, referred to as the {\it star-product}, which replaces the classical
pointwise multiplication of functions on phase-space. It was shown that the 
full machinery of quantum mechanics can be reformulated by working with the
algebra of functions on phase-space, whereby the algebra-multiplication is 
provided by the star-product. Since the star-product, in the classical limit,
reduces to the pointwise product, the former may be considered a "quantum
deformation" of the latter. Clearly this deformation-theory approach to
quantization\refmark{8} is very much in the spirit of modern non-commutative
geometry: the transition from classical to quantum phase-space is
achieved by deforming the algebra of functions on the space under
consideration.
\par
Moyal's phase space formulation of quantum meachanics makes use of
the so-called symbol calculus\Ref\symbo{F.A.Berezin, Sov.Phys.Usp.
23 (1980) 1981;\nextline
L.H\"ormander, Comm. Pure and Applied Math. 32 (1979) 359}
which associates, in a one-to-one manner, ordinary functions to the
operators on some Hilbert space. In this way the observables and the density
operators of the standard Hilbert space formulation of quantum mechanics are
turned into functions on phase-space; they are called the "symbols" of the 
respective operators. Operator products correspond to star-products of symbols
then, and commutators of operators go over into the Moyal bracket, which is the
commutator with respect to star-multiplication. The Moyal bracket is a
deformation\refmark{7} of the classical Poisson bracket, to which it reduces in the
limit~$\hbar\rightarrow 0$. Moyal has studied the quantum dynamcs of scalar
pseudo-densities (Wigner functions\refmark{9}) in this language. In doing so he 
introduced a quantum deformed version of the classical hamiltonian vector
field. It is the purpose of the present paper to investigate more general
geometric objects and operations such as vectors, forms, Lie derivatives,
etc., in this framework. In a previous paper\Ref\clas{E.Gozzi, M.Reuter,
W.D.Thacker, Phys.Rev.D 40 (1989) 3363; ibid.~D46 (1992) 757} we have
reformulated the classical exterior calculus on symplectic manifolds
(Cartan calculus) in a hamiltonian language. This means that operations such
as Lie derivatives, exterior derivatives, contractions, etc. were expressed in
terms of a novel type of Poisson bracket defined for functions on an extended
phase-space. The extended phase-space is a supermanifold\Ref\dew{B.DeWitt, 
"{\it Supermanifolds}",~Cambridge University Press, ~1984}
which is closely related to the (co)-tangent bundle over the standard
phase-space. In this way, scalar functions on the extended phase-space
are equivalent to tensors on the standard phase-space. However, as discussed
above, we know how to deform the algebra of (scalar) functions on any
phase-space, therefore we should arrive at a kind of {\it "quantum exterior
calculus"} if we apply the Moyal deformation not to the standard phase-space,
but rather to the extended one. In order to implement this program we
first review in section 2 the relevant material on the Moyal deformation. 
Then, in section 3, we introduce the extended phase-space and describe the
classical Cartan calculus in terms of the associated extended Poisson
bracket structure. Finally, in sections 4 through 7, we study the
deformed calculus resulting from the Moyal deformation of the extended
Poisson bracket structure.
\chapter{SYMBOL CALCULUS AND WEYL-WIGNER-MOYAL FORMALISM}
The basic idea behind the "symbol calculus"\refmark{7-10} is to set up
a linear one-to-one map between the operators\foot{In this section the
caret~${\widehat{(\cdot)}}$~is used to denote operators.}~${\widehat A}$,
~${\widehat B},\cdots$~on some Hilbert space ~${\cal V}$~ and the 
~complex-valued functions~$A,B,\cdots\in Fun({\cal M})$ defined on an appropriate
finite-dimensional manifold ~${\cal M}$. The operator ~${\widehat A}$~is
uniquely represented by the function ~$A$~ which is called the symbol
of ~${\widehat A}$. For the "symbol map" relating the function ~$A$~to the
operator ~${\widehat A}$~ we write ~$A=symb({\widehat A})$. It has
a well-defined inverse~${\widehat A}=symb^{-1}(A)$. The space of symbols,
~$Fun({\cal M})$, is equipped with the so called "star-product"
~$\ast$~which implements the operator multiplication at the level of symbols.
It is defined by the requirement that the symbol map is an algebra 
homomorphism, \ie,~that
$$ symb({\widehat A}{\widehat B})=symb({\widehat A})\ast
symb({\widehat B})\eqn\francesco$$
for any pair of operators ~${\widehat A}$~and ~${\widehat B}$. Since operator
multiplication is non-commutative in general, but associative, the
same is also true for the star-multplication:
$$A\ast B\ne B\ast A\eqn\gianni$$
$$A\ast(B\ast C)=(A\ast B)\ast C\eqn\daniela$$
As we shall see, the star-product may be considered a deformation\refmark{8}
of the ordinary pointwise product of functions. Here "deformation" is meant
in the sense of ref.[13]\REF\vech{M.Gerstenhaber, Ann.of Math. 79 (1964) 59}
~to which we refer the reader for further details.
\par
Let us now be more specific and let us us assume that the
Hilbert space ~${\cal V}$~is the state space of an arbitrary
quantum mechanical system with ~$N$~degrees of freedom, and that
the manfold~${\cal M}={\cal M}_{2N}$~is the $2N$-dimensional classical phase-space
pertaining to this system. Then the quantum mechanical operator
~${\widehat A}$~is represented by a function~$A=A(\phi)$,~where 
~$\phi^{a}=(p^{1},\cdots,p^{N},q^{1},\cdots, q^{N}),~a=1,\cdots,2N$~are
canonical coordinates on the phase-space ~${\cal M}_{2N}$. For the sake of
simplicity we assume that canonical coordinates can be introduced globally.
This implies that the symplectic two form\Ref\clas{V.I.Arnold, "{\it 
Mathematical Methods of Classical Mechanics}",~Springer, New York, 1978;\nextline
R.Abraham and J.Marsden, "{\it Foundations of Mechanics}",~Benjamin, New York,~
1978} on ~${\cal M}_{2N}$,~$\omega={1\over 2}\omega_{ab}d\phi^{a}\wedge
d\phi^{b}$,~has constant components:
$$\omega_{ab}=\pmatrix{0 & I_{N} \cr
-I_{N}& 0 \cr}\eqn\vitto$$
The inverse matrix, denoted by ~$\omega^{ab}$,~reads
$$\omega^{ab}=\pmatrix{0 & -I_{N} \cr
I_{N} & 0 \cr}\eqn\lidia$$
Using~$\omega^{ab}$~we define the Poisson bracket for any pair of functions
~$A,B\in Fun({\cal M}_{2N})$:
$$\{A,B\}_{pb}(\phi)\equiv
\partial_{a}A(\phi)\omega^{ab}\partial_{b}B(\phi)\eqn\pepo$$
Here ~$\partial_{a}\equiv{\partial\over\partial\phi^{a}}$.
\par
There exists a variety of possibilities of associating an operator\break
~${\widehat A}({\widehat p},{\widehat q})=symb^{-1}\bigl(A(p,q)\bigr)$
to the function ~$A(p,q)$. Typically, different definitions of the symbol-map
correspond to different operator ordering prescriptions\refmark{10}. 
In the following we shall mainly work with the Weyl symbol\refmark{9}
which has the property that if ~$A(p,q)$~ is a polynomial in ~$p$~and
~$q$, the operator ~${\widehat A}({\widehat p},{\widehat q})$~is the 
symmetrically ordered polynomial in ~${\widehat p}$~and
~${\widehat q}$,~\eg,~$symb^{-1}(pq)={1\over 2}({\widehat p}
{\widehat q}+{\widehat q}{\widehat p})$. The Weyl symbol ~$A(\phi^{a})$~
of the operator~${\widehat A}$~is given by\refmark{10}
\Ref\litt{R.G.Littlejohn, Phys.Rep.138 (1986) 193}
$$A(\phi^{a})=\int{d^{2N}\phi_{0}\over (2\pi\hbar)^{N}}~exp\bigl[{i\over
\hbar}\phi_{0}^{a}\omega_{ab}\phi^{b}\bigr]Tr\bigl[{\widehat T}(\phi_{0})
{\widehat A}\bigr]\eqn\paver$$
where
$${\widehat T}(\phi_{0})=exp\bigl[{i\over \hbar}\phi^{a}\omega_{ab}
{\widehat \phi}^{b}_{0}\bigr]\equiv exp\bigl[{i\over\hbar}(p_{0}{\widehat q}
-q_{0}{\widehat p})\bigr]$$
The inverse map reads
$${\widehat A}=\int {d^{2N}\phi~d^{2N}\phi_{0}\over (2\pi\hbar)^{2N}}
~A(\phi)~exp\bigl[{i\over\hbar}\phi^{a}\omega_{ab}\phi_{0}^{b}\bigr]
{\widehat T}(\phi_{0})\eqn\spallucci$$
Eq.\spallucci~is due to Weyl\refmark{9}. It expresses the fact that the
operators~${\widehat T}(\phi_{0})$~form a complete and orthogonal
(with respect to the Hilbert-Schmidt inner product) set of operators in terms
of which any operator ~${\widehat A}$~ can be expanded\refmark{15}.
\par
A particularly important class of operators are the density
operators~${\widehat\varrho}$. Their symbols ~$\varrho=symb({\widehat
\varrho})$~are given by eq.~\paver~for ~${\widehat A}={\widehat\varrho}$.
In particular, for pure states ~${\widehat\varrho}=\ket{\psi}\bra{\psi}$~
one obtains the Wigner function\refmark{9}
$$\varrho(p,q)=\int d^{N}x~exp\bigl[-{i\over\hbar}px\bigr]
\psi(q+{1\over 2}x)\psi^{\ast}(q-{1\over 2}x)\eqn\furlan$$
The symbol ~$\varrho(\phi^{a})$~is the quantum mechanical analogue
of the classical probability density~$\varrho_{cl}(\phi^{a})$~used in
classical statistical mechanics. However, differently than in 
classical mechanics, the quantum symbol~$\varrho(\phi)$~ is not positive
definite and is therefore referred to as a "pseudodensity". The usual
positive definite quantum mechanical distributions over position or
momentum space, respectively, are recovered as
$$\vert\psi(q)\vert^{2}=\int{d^{N}p\over (2\pi\hbar)^{N}}~\varrho(p,q)$$
$$\vert{\widetilde\psi}(p)\vert^{2}=\int{d^{N}q\over (2\pi\hbar)^{N}}~\varrho
(q,p)$$
Similarly, the expectation value of any observable~${\widehat{\cal O}}$~is 
given by
$$<\psi\vert{\widehat{\cal O}}\vert\psi>=\int{d^{2N}\phi\over (2\pi\hbar)^{N}}
~\varrho(\phi){\cal O}(\phi)$$
In this way quantum mechanics can be formulated in a "classically-looking"
manner involving only c-number functions on~${\cal M}_{2N}$.
\par
The star product which makes the algebra of Weyl symbols isomorphic to
the operator algebra is given by
$$\eqalign{\bigl(A\ast B\bigr)(\phi) \ = & \ A(\phi)~exp\bigl[i{\hbar\over 2}
\stackrel{\leftarrow}{\partial_{a}}\omega^{ab}
\stackrel{\rightarrow}{\partial_{b}}\bigr]B(\phi) \cr
\ \equiv & \ exp\bigl[i{\hbar\over 2}\omega^{ab}
\stackrel{1}{\partial_{a}}\stackrel{2}{\partial_{b}}\bigr]
A(\phi_{1})~B(\phi_{2})\vert_{\phi_{1}=\phi_{2}=\phi}
\cr}\eqn\vittos$$
with~$\stackrel{1,2}{\partial_{a}}={\partial\over\partial
\phi^{a}_{1,2}}$,~or more explicitly
$$\eqalign{(A\ast B)(\phi) \ = & \ \sum_{m=0}^{\infty}{1\over
m!}({i\hbar\over 2})^{m}\omega^{a_{1}b_{1}}\cdots\omega^{a_{m}b_{m}}
(\partial_{a_{1}}\cdots\partial_{a_{m}}A)(\partial_{b_{1}}\cdots\partial
_{b_{m}}B)\cr
\ = & \ A(\phi)B(\phi)+O(\hbar)\cr}\eqn\vitti$$
We see that to lowest order in ~$\hbar$~the star-product of two
functions reduces to the ordinary pointwise product. For
non-zero values of ~$\hbar$~ this multiplication is "deformed"
in such a way that the resulting ~$\ast$-product remains associative
but non-commutative in general.
\par
The Moyal bracket\refmark{7} of two symbols ~$A$,~$B\in Fun
({\cal M}_{2N})$~is defined as their commutator (up to a factor of
~$i\hbar$) with respect to star-multiplication:
$$\eqalign{\bigl\{A,B\bigr\}_{mb} \ = & \ {1\over i\hbar}\bigl(A\ast B-B\ast
A\bigr)\cr
\ = & \ symb\bigl({1\over i\hbar}[{\widehat A},{\widehat B}]
\bigr)\cr}\eqn\vittu$$
Using ~\vittos~this can be written as
$$\eqalign{\bigl\{A,B\bigr\}_{mb} \ = & \ A(\phi)~{2\over\hbar}sin\bigl[
{\hbar\over 2}\stackrel{\leftarrow}{\partial_{a}}\omega^{ab}\stackrel
{\rightarrow}{\partial_{b}}\bigr]B(\phi)\cr
\ = & \ \bigl\{A,B\bigr\}_{pb}+O(\hbar^{2})\cr}\eqn\vittus$$
In the classical limit ~($\hbar\rightarrow 0$)~the Moyal bracket reduces
to the classical Poisson bracket. The Moyal bracket ~$\bigl\{\cdot,\cdot
\bigr\}_{mb}$~is a "deformation" of ~$\bigl\{\cdot,\cdot\bigr\}_{pb}$~
which preserves two important properties of the Poisson bracket:
\nextline
({\bf i}) The Moyal bracket obeys the Jacobi identity.\nextline
({\bf ii})~For every ~$A\in Fun({\cal M}_{2N})$~ the operation
 ~$\bigl\{A,\cdot\bigr\}_{mb}$~is a derivation on the algebra
~$\bigl(Fun({\cal M}_{2N}),\ast\bigr)$,~\ie,~
it obeys a Leibniz rule\foot{Recall that, for any algebra with elements
~$A,B,\cdots$~ and a product $\circ$, a derivation ~${\cal D}$~ has
the property ~${\cal D}(A\circ B)=({\cal D}A)\circ B+A\circ 
({\cal D}B)$. In the present case, $A\circ B\equiv\bigl\{A,B\bigr\}_{mb}$.}
of the form
$$\bigl\{A,B_{1}\ast B_{2}\bigr\}_{mb}=\bigl\{A,B_{1}\bigr\}\ast
B_{2}+B_{1}\ast\bigl\{A,B_{2}\bigr\}_{mb}\eqn\gaetano$$
It is also a well known fact\refmark{8} that all derivations
~${\cal D}$~ of the Moyal bracket  are "inner derivation",~\ie,~for any
~${\cal D}$~ one can find an element ~$X\in Fun({\cal M}_{2N})$~ such that
~${\cal D}A=\bigl\{X,A\bigr\}_{mb}$. An analogous statement holds true for
the commutator algebra, but not for the classical Poisson-bracket
algebra. This difference of the algebraic properties of the Moyal and
Poisson brackets is also at the heart of the Groenwald-Van Hove obstruction to
quantization\Ref\ogre{H.J.Groenwald, Physica 12 (1946) 405;\nextline
L.Van Hove, Mem.Acad. R.Belg. 26 (1951) 26}. The standard correspondence 
rules of quantum mechanics as postulated by Dirac~try to associate operators
~${\widehat A}$~to phase functions ~$A(\phi)$~ in such a way that the
{\it Poisson} bracket algebra is matched by the operator algebra.
In view of the above discussion, which shows that actually it is
the Moyal bracket algebra which is equivalent to the operatorial one,
it is clear that the Dirac correspondence can be implemented only for the
very narrow class of observables for which the higher derivatives of the RHS 
of eq.\vittus~are ineffective,~\ie,~for functions at most quadratic in
~${\phi}^{a}$.
\par
The symbol calculus suggests that the process of "quantization" can be
understood as a smooth deformation of the algebra of classical observables
$\bigl(\{\cdot,\cdot\}_{pb}\rightarrow\{\cdot,\cdot\}_{mb}\bigr)$~rather
than as a radical change in the nature of the observables~$\bigl($~
c-numbers~$\mapsto$~operators~$\bigr)$. This point of view has been advocated in
refs.[8] and in Moyal's original paper\refmark{7}~where also the time evolution
of the pseudodensities~$\varrho(\phi,t)$~ has been studied. At the operatorial
level we have von Neumann's equation
$$i\hbar~\partial_{t}{\widehat\varrho}=-\bigl[{\widehat\varrho},{\widehat
H}\bigr]$$
which goes, via the symbol map, in
$$\partial_{t}\varrho(\phi^{a},t)=-\bigl\{\varrho,H\bigr\}_{mb}\eqn\tullio$$
where ~$H(\phi)$~is the symbol of the hamiltonian operator
~${\widehat H}$. For pure states with ~${\widehat\varrho}=\ket{\psi}\bra
{\psi}$~this equation is equivalent to the Schr\"odinger equation for
~$\ket{\psi}$. In the classical limit eq.~\tullio~ becomes the well-known
Liouville equation for a (positive-definite) probability density~$\varrho$~:
$$\eqalign{\partial_{t}\varrho(\phi^{a},t) \ = & \ 
-\bigl\{\varrho,H\bigr\}_{pb}\cr
\ \equiv & \ -h^{a}(\phi)\partial_{a}\varrho\cr
\ \equiv & \ -l_{h}\varrho\cr}\eqn\nello$$
In the second line of eq.~\nello~we used the components
$$h^{a}(\phi)\equiv\omega^{ab}\partial_{a}H(\phi)\eqn\euro$$
of the hamiltonian vector field\refmark{14} ~$h\equiv h^{a}\partial_{a}$, which
coincides with the Lie derivative ~$l_{h}$~when acting on scalars
(zero-forms)~$\varrho(\phi)$. Comparing eqs.~\tullio~and ~\nello~ we may
say that the Moyal bracket gives rise to the notion of a {\it quantum
deformed hamiltonian vector field} or, equivalently, of a quantum deformed 
Lie derivative for zero-forms.
\par
In classical mechanics it is well known\refmark{14} how to generalize
eq.~\nello~ to higher p-form valued "densities" of the type
$$\varrho={1\over p!}~\varrho_{a_{1}\cdots a_{p}}(\phi)~d\phi^{a_{1}}\wedge
\cdots\wedge d\phi^{a_{p}}\eqn\franco$$
They are pulled along the hamiltonian flow according to the equation of motion
$$\partial_{t}\varrho=-l_{h}\varrho\eqn\ghiro$$
where now~$l_{h}$~is the Lie derivative appropriate for p-forms. So far
no quantum mechanical analogue of eq.~\ghiro~has been constructed along 
the lines of Moyal. It is exactly this problem which we shall address in
the following sections. As mentioned in the introduction, our
strategy is to introduce an extended phase-space, denoted by 
~${\cal M}_{8N}$, such that scalars on ~${\cal M}_{8N}$~represent 
antisymmetric tensors on the standard phase-space ~${\cal M}_{2N}$,
and to apply the Moyal deformation to the extended phase-space.
\chapter{FROM ORDINARY PHASE-SPACE TO EXTENDED PHASE-SPACE}
In this section we describe how the conventional exterior calculus on 
phase-space can be reformulated in a hamiltonian language which lends
itself to a deformation \`a la Moyal. In the present section we introduce the
relevant classical structures. Their Moyal deformation will be
discussed later on.
\par
Let us suppose we are given a ~$2N$-dimensional symplectic manifold
~${\cal M}_{2N}$~ endowed with a closed non-degenerate
two form~$\omega$. For simplicity we assume again that we can introduce
canonical coordinates globally so that the components ~$\omega_{ab}$
~are given by~\vitto. Furthermore we pick some Hamiltonian
~$H\in Fun({\cal M}_{2N})$. It gives rise to the vector field
~$h^{a}$~of eq.~\euro~ in terms of which Hamilton's equations
read
$${\dot\phi}^{a}(t)=h^{a}(\phi(t))\eqn\pino$$
In the following we consider ~$h^{a}$~in its role as the generator
of symplectic diffeomorphisms (canonical transformations). Under the
transformation
$$\delta\phi^{a}=-h^{a}(\phi)\eqn\paolo$$
the components of any tensors change according to
$$\delta T^{a_{1}a_{2}\cdots}_{b_{1}b_{2}\cdots}=l_{h}T^{a_{1}a_{2}\cdots}
_{b_{1}b_{2}\cdots}\eqn\daniele$$
where
$$l_{h}T^{a\cdots}_{b\cdots}=h^{c}\partial_{c}T^{a\cdots}_{b\cdots}
+\partial_{b}h^{c}T^{a\cdots}_{c\cdots}-\partial_{c}h^{a}
T^{c\cdots}_{b\cdots}+\cdots\eqn\ursula$$
is the classical Lie-derivative\refmark{14}.
As time evolution in classical mechanics is a special symplectic diffeomorphism,
the time-evolution of the p-form density~\franco~is given by
$$\partial_{t}\varrho_{a_{1}\cdots a_{p}}(\phi,t)=-l_{h}\varrho
_{a_{1}\cdots a_{p}}(\phi,t)\eqn\viviana$$
As we mentioned already, for ~$p=0$~eq.~\viviana~coincides with Liouville's
equation~\nello. In this case we were able to give a hamiltonian interpretation
to the RHS of the evolution equation: it was the Poisson bracket of
~$H$~with~$\varrho$. We shall now introduce the extended phase-space
~${\cal M}_{8N}$~in such a manner that, even for ~$p>0$,
the RHS of eqn.~\viviana~ can be expressed as a generalized Poisson
bracket.
\par
The extended phase-space ~${\cal M}_{8N}$~ is a ~$4N+4N$~dimensional
supermanifold\refmark{12} with ~$4N$~bosonic and ~$4N$~fermionic
dimensions. It is coordinatized by the $8N$-tuples\break
~$\bigl(\phi^{a},\lambda_{a},c^{a},{\bar c}_{a}\bigr)~,~a=1\cdots 2N$.
Here ~$\phi^{a}$~are coordinates on the standard phase-space~${\cal M}_{2N}$
~which we now identify with the hypersurface in ~${\cal M}_{8N}$~on
which ~$\lambda_{a}=0$~and~$c^{a}=0={\bar c}_{a}$. The ~$\lambda_{a}$'s
are additional bosonic variables, and the ~$c^{a}$'s and ~${\bar c}_{a}$'s
are anticommuting Grassmann numbers. As indicated by the positioning of
the indices,~$\lambda_{a}$~and ~${\bar c}_{a}$~are assumed to transform,
under a diffeomorphism on ~${\cal M}_{2N}$, like 
the derivatives~$\partial_{a}$, while ~$c^{a}$~transform like the coordinate 
differentials ~$d\phi^{a}$.
Let us define the extended Poisson bracket ~$(epb)$~structure on
~${\cal M}_{8N}$~as follows
$$\eqalign{\ \bigl\{ & \ \phi^{a},\lambda_{b}\bigr\}_{epb}=\delta^{a}_{b}~~,
~~\bigl\{\phi^{a},\phi^{b}\bigr\}_{epb}=0=\bigl\{\lambda_{a},\lambda_{b}
\bigr\}\cr
\ \bigl\{ & \ c^{a},{\bar c}_{b}\bigr\}_{epb}=-i\delta^{a}_{b}~~,~~all~others
=0\cr}\eqn\aste$$
With respect to the ~$epb$-structure, the auxiliary variables ~$\lambda_{a}$
~can be thought of as "momenta" conjugate to the ~$\phi^{a}$'s. Note also 
that the ~$\phi^{a}$'s have vanishing extended Poisson brackets among
themselves, whereas their conventional Poisson bracket on ~${\cal M}_{2N}$~
is different from zero:
$$\bigl\{\phi^{a},\phi^{b}\bigr\}_{pb}=\omega^{ab}\eqn\bala$$
Eq.~\aste~implies the following bracket for ~$A,B\in Fun({\cal M}_{8N})$,~\ie
,~for functions ~$A=A(\lambda_{a},\phi^{a},{\bar c}_{a},c^{a}),\cdots$
$$\bigl\{A,B\bigr\}_{epb}= A\bigl[{\stackrel{\leftarrow}{\partial}\over
\partial \phi^{a}}{\stackrel{\rightarrow}{\partial}\over\partial\lambda_{a}}
-{\stackrel{\leftarrow}{\partial}\over\partial\lambda_{a}}{\stackrel
{\rightarrow}{\partial}\over\partial\phi^{a}}-i\bigl(
{\stackrel{\leftarrow}{\partial}\over\partial {\bar c}_{a}}{\stackrel
{\rightarrow}{\partial}\over\partial c^{a}}+
{\stackrel{\leftarrow}{\partial}\over\partial c^{a}}
{\stackrel{\rightarrow}{\partial}\over\partial {\bar c}_{a}}\bigr)
\bigr]B\eqn\marm$$
(Note that this is a ~${\bf Z_{2}}$-graded bracket whose symmetry
character depends on whether ~$A$~ and ~$B$~ are even or odd
elements of the Grassmann algebra.) The $epb$-structure~\aste~was
first introduced in refs.[11] where we gave a path-integral 
representation of classical hamiltonian mechanics. In particular, 
it was shown that the Grassmann variables ~${\bar c}_{a}$~form
a basis in the tangent space~$T_{\phi}{\cal M}_{2N}$~ and that,
similarly, the ~$c^{a}$'s form a basis in the cotangent space
~$T^{\ast}_{\phi}{\cal M}_{2N}$~thus playing the role of the
differentials ~$d\phi^{a}$. This fact can be exploited as follows.
Let us assume we are given an arbitrary, completely antisymmetric tensor 
field on ~${\cal M}_{2N}$:
$$T=T^{b_{1}\cdots b_{q}}_{a_{1}\cdots a_{p}}(\phi)~\partial_{b_{1}}\wedge
\cdots\wedge\partial_{b_{q}}~d\phi^{a_{1}}\wedge\cdots\wedge d\phi^{a_{p}}
\eqn\piero$$
From ~$T$~ we can construct the following function
~${\widehat T}\in Fun({\cal M}_{8N})$:
$${\widehat T}=T^{b_{1}\cdots b_{q}}_{a_{1}\cdots a_{p}}(\phi)~
{\bar c}_{b_{1}}\cdots {\bar c}_{b_{q}}~c^{a_{1}}\cdots c^{a_{p}}\eqn\liana$$
Under diffeomorphisms on ~${\cal M}_{2N}$~ the function ~${\widehat T}$~
transforms as a scalar. Here and in the following the caret~
${\widehat{(\cdot)}}$~does not indicate operators, but rather that
~$\partial_{a}$~ and ~$d\phi^{a}$~ have been replaced by ~${\bar c}_{a}$
and ~$c_{a}$, respectively. Sometimes we refer to this substitution
as the "hat map". For example, the p-form of eq.~\franco~ becomes
$${\widehat\varrho}={1\over p!}~\varrho_{a_{1}\cdots a_{p}}(\phi)
~c^{a_{1}}\cdots c^{a_{p}}\eqn\cabibbo$$
We mentioned already that we would like to express the RHS of eq.~\viviana~
like a Poisson bracket as it was done in eq.~\nello~for zero forms. To this
end we try to find a "super-Hamiltonian"~${\widetilde{\cal H}}\in Fun ({\cal M}
_{8N})$~with the following two properties
$$\bigl\{{\widetilde{\cal H}},\varrho\bigr\}_{epb}=\bigl\{H,\varrho\bigr\}_{pb}
\equiv -l_{h}\varrho\eqn\maiani$$
$$\bigl\{{\widetilde{\cal H}},\varrho_{a_{1}\cdots a_{p}}(\phi)c^{a_{1}}\cdots
c^{a_{p}}\bigr\}_{epb}=-\bigl(l_{h}\varrho_{a_{1}\cdots a_{p}}\bigr)c^{a_{1}}\cdots
c^{a_{p}}\eqn\lusignoli$$
Eq.~\maiani~ guarantees that for zero-form~$\varrho=\varrho(\phi)$~the
dynamics given by the new Hamiltonian ~${\widetilde{\cal H}}$~together
with the extended Poisson bracket coincides with the one obtained from the
standard Hamiltonian together with the standard Poisson bracket. Eq.
~\lusignoli~ generalizes eq.~\maiani~for higher forms. It allows us to
rewrite eq.~\viviana~in hamiltonian form:
$$\partial_{t}{\widehat\varrho}=-\bigl\{{\widehat\varrho},{\widetilde{\cal H}}
\bigr\}_{epb}\eqn\martinelli$$
In ref.[11] we showed that the solution to eqs.~\maiani~and ~\lusignoli~is
provided by the following super-Hamiltonian
$${\widetilde{\cal H}}={\widetilde{\cal H}}_{B}+
{\widetilde{\cal H}}_{F}\eqn\pari$$
where
$${\widetilde{\cal H}}_{B}=\lambda_{a}h^{a}(\phi)\eqn\paris$$
and
$${\widetilde{\cal H}}_{F}=i{\bar c}_{a}\partial_{b}h^{a}(\phi)c^{b}\eqn\parisi$$
The super-Hamiltonian ~${\widetilde{\cal H}}$~has vanishing~$ep$-bracket with
the following conserved charges:
$$\eqalign{Q \ = & \ ic^{a}\lambda_{a}\cr
{\bar Q} \ = & \ i{\bar c}_{a}\omega^{ab}\lambda_{b}\cr
Q_{g} \ = & \ c^{a}{\bar c}_{a}\cr
K \ = & \ {1\over 2}\omega_{ab}c^{a}c^{b}\cr
{\bar K} \ = & \ {1\over 2}\omega^{ab}{\bar c}_{a}
{\bar c}_{b}\cr}\eqn\valentina$$
Under the extended Poisson bracket they form a closed algebra isomorphic
to ~$ISp(2)$, whose inhomogeneous part is generated\refmark{11} by the BRS operator 
~$Q$~ and the anti-BRS operator ~${\bar Q}$. It is interesting that
~${\widetilde{\cal H}}$~is a pure BRS-variation,
$${\widetilde{\cal H}}=i\bigl\{Q\bigl\{{\bar Q},H\bigr\}_{epb}\bigr\}_{epb}
\eqn\claudia$$
which is  typical of topological field theories\Ref\wit{E.Witten, Com.Math.
Phys.117 (1988) 353;~ibid. 118 (1988) 411;\nextline
D.Birmingham, M.Blau, M.Rakowski, G.Thompson, Phys.Rep.209 (1991) 130}. 
The use of it in this
context has been explored in ref.[18]\REF\nos{E.Gozzi, M.Reuter, 
Phys.Lett. B 240(1990) 137}. Furthermore, ~${\widetilde{\cal H}}$~possesses 
a N=2 supersymmetry\Ref\sus{E.Gozzi, M.Reuter, Phys.Lett.~B 233 (1989) 383;
ibid. B 238 (1990) 451}
justifying the term "super-Hamiltonian" for
~${\widetilde{\cal H}}$. However the SUSY will play no important role in the 
following.
\par
The five conserved charges ~\valentina~ are the essential tool in 
reformulating the classical Cartan calculus in hamiltonian form. To
illustrate this point, consider the following tensors  and their
counterparts in ~$Fun({\cal M}_{8N})$:
$$\eqalign{v=v^{a}\partial_{a}\  \longmapsto & \ {\widehat v}=v^{a}
{\bar c}_{a}\cr
\alpha=\alpha_{a}d\phi^{a} \ \longmapsto & \ {\widehat\alpha}=\alpha_{a}
c^{a}\cr
F^{(p)}={1\over p!}F_{a_{1}\cdots a_{p}}~d\phi^{a_{1}}\wedge\cdots\wedge d\phi^{a_{p}}
\ \longmapsto & \ {\widehat F}^{(p)}={1\over p!}F_{a_{1}\cdots a_{p}}~c^{a_{1}}\cdots
c^{a_{p}}\cr
V^{(p)}={1\over p!}V^{a_{1}\cdots a_{p}}~\partial_{a_{1}}\wedge\cdots\wedge\partial_{a_{p}}
\ \longmapsto & \ {\widehat V}^{(p)}={1\over p!}V^{a_{1}\cdots a_{p}}~{\bar c}_{a_{1}}
\cdots {\bar c}_{a_{p}}\cr}\eqn\maria$$
All coefficients~$v^{a},\alpha_{a},F_{a_{1}\cdots a_{p}}, etc.$~
appearing in the above formulas are functions of ~$\phi$.
By this {\it "hat map"}~~${\widehat\cdot}$,~~the exterior derivative of p-forms
~$F^{(p)}$~goes over into the ~$ep$-bracket with the BRS charge ~${Q}$:
$$\bigl(dF^{(p)}\bigr)^{\wedge}=i\bigl\{Q,
{\widehat F}^{(p)}\bigr\}_{epb}\eqn\antonia$$
The exterior co-derivative of p-vectors~$V^{(p)}$~is given by the ~$ep$-bracket
with the anti-BRS operator~${\bar Q}$
$$\bigl({\bar d}V^{(p)}\bigr)^{\wedge}=i\bigl\{
{\bar Q},{\widehat V}^{(p)}\bigr\}_{epb}\eqn\madona$$
$${\bar d}V^{(p)}=\omega^{ab}\partial_{b}V^{a_{1}\cdots a_{p}}~\partial_{a}\wedge
\partial_{a_{1}}\cdots\wedge\partial_{a_{p}}\eqn\matone$$
In symplectic geometry vectors and forms can be related by contraction with
~$\omega_{ab}$~or ~$\omega^{ab}$. This operation is realized\refmark{11,14}
as the ~$ep$-bracket with ~$K$~and ~${\bar K}$:
$$\eqalign{\bigl(v^{\flat}\bigr)^{\wedge} \ = & \ i\bigl\{K,{\widehat v}\bigr\}
_{epb}~~,~~\bigl(v^{\flat}\bigr)_{a}\equiv \omega_{ac}v^{c}\cr
\bigl(\alpha^{\sharp}\bigr)^{\wedge} \ = & \ i\bigl\{{\bar K},{\widehat 
\alpha}\bigr\}_{epb}~~,~~\bigl(\alpha^{\sharp}\bigr)^{a}\equiv\omega^{ac}
\alpha_{c}\cr}\eqn\bonora$$
The contractions with vectors and 1-forms translates into the following 
brackets
$$\eqalign{\bigl(i(v)F^{(p)}\bigr)^{\wedge} \ = & \ i\bigr\{{\widehat v}
,{\widehat F}^{(p)}\bigr\}_{epb}\cr
\bigl(i({\alpha}) V^{(p)}\bigr)^{\wedge} \ = & \ i\bigl\{{\widehat\alpha},
{\widehat V}^{(p)}\bigr\}_{epb}\cr}\eqn\amati$$
The Lie derivative along the hamiltonian vector field is
$$\bigl(l_{h}T\bigr)^{\wedge}=-\bigl\{{\widetilde{\cal H}},
{\widehat T}\bigr\}_{epb}\eqn\jengo$$
where~$T$~ can be any antisymmetric tensor. The ~$epb$-representation
of the various {\it classical}-tensor manipulations are summarized
in table 1.
\chapter{MOYAL DEFORMATION ON EXTENDED PHASE-SPACE}
In the previous section we realized the classical Cartan calculus in terms 
of ~$ep$-brackets on the extended phase-space ~${\cal M}_{8N}$. Let us now
try to deform the extended Poisson bracket to an extended
Moyal bracket. Following Berezin\refmark{10},~we define the
{\it extended star product}\foot{To keep  with
the supersymmetry jargon we should call
this product the ~"{\it super-star product}".} on ~$Fun({\cal M}_{8N})$~as
$$A\ast_{e}B\equiv A~exp\bigl[{i\over 2}\bigl({\stackrel{\leftarrow}
{\partial}\over\partial\phi^{a}}{\stackrel{\rightarrow}{\partial}
\over\partial\lambda_{a}}-{\stackrel{\leftarrow}{\partial}\over
\partial\lambda_{a}}{\stackrel{\rightarrow}{\partial}\over\partial\phi^{a}}
\bigr)+{\stackrel{\leftarrow}{\partial}\over\partial c^{a}}{\stackrel
{\rightarrow}{\partial}\over\partial {\bar c}_{a}}\bigr]B\eqn\pordenone$$
where ~$A(\lambda,\phi,{\bar c},c)$,~etc.
The extended Moyal bracket is introduced as the graded commutators with respect
to~$\ast_{e}$-multiplication:
$$\bigl\{A,B\bigr\}_{emb}={1\over i}\bigl[A\ast_{e}B- (-)^{[A][B]}B\ast_{e}
A\bigr]\eqn\verona$$
Here~$[A]=0,1$~denotes the grading of ~$A$~. The sign factor on the RHS of
~\verona~guarantees that the ~$em$-bracket has the same symmetry
properties as the graded commutator:
$$\bigl\{A,B\bigr\}_{emb}=-(-1)^{[A][B]}\bigl\{B,A\bigr\}_{emb}$$
It can be checked that the bracket~\verona~obeys the graded Jacobi
identity and that ~$\bigl\{A,{\cdot}\bigr\}_{emb}$~is a graded derivation
of the algebra~$\bigl(Fun({\cal M}_{8N}),\ast_{e}\bigr)$~for any
~$A\in Fun({\cal M}_{8N})$:
$$\bigl\{A,B_{1}\ast_{e}B_{2}\bigr\}_{emb}=\bigl\{A,B_{1}\bigr\}
\ast_{e}B_{2}+(-1)^{[A][B_{1}]}B_{1}\ast_{e}\bigl\{A,B_{2}
\bigr\}_{emb}\eqn\mantova$$
This is a consequence of the associativity of the extended star product.
Eq.~\mantova~is very important for our purposes because we would like to 
rewrite the derivations~$d,i_{v},l_{h},etc.,$~as extended Moyal brackets.
Comparing~\pordenone~to~\vittos~we see that the transition from standard
phase-space to extended phase-space entails the replacements
~${\partial\over\partial q}\rightarrow {\partial\over\partial\phi}$,
~${\partial\over\partial p}\rightarrow {\partial\over\partial\lambda}$~
and the addition of the Grassmannian piece. For a reason which will
become clear shortly, we have set~$\hbar=1$~for the deformation parameter
in eq.~\pordenone. Reinstating ~$\hbar$~and letting ~$\hbar\rightarrow 0$,
the extended Moyal bracket~\verona~reduces to the extended Poisson bracket
~\marm. The fundamental ~$em$-brackets among ~$\phi^{a},\lambda_{a},
c^{a}$~and~${\bar c}_{a}$~coincide with the respective ~$ep$-brackets
given in eq.~\aste~because in this case the higher derivative terms
vanish. In practical calculations, involving the extended star product,
the following alternative representation has proven helpful:
$$\eqalign{(A\ast_{e} B)(\phi,\lambda,{\bar c}, c) \ = & \ A 
\bigl(\lambda_{a}-{i\over 2}
{\partial\over\partial{\tilde\phi}^{a}},\phi^{a}+{i\over 2}{\partial\over
\partial{\tilde\lambda}_{a}},{\bar c}_{a},c^{a}+{\partial\over\partial
{\tilde{\bar c}}}\bigr)B({\tilde\lambda},{\tilde\phi},{\tilde{\bar c}},c)
\vert_{{\tilde\lambda}=\lambda\atop {{\tilde\phi}=\phi\atop{\tilde{\bar c}}
={\bar c}}}\cr
\ = & \ A({\tilde\lambda},{\tilde\phi},{\bar c},{\tilde c})B\bigl(\lambda_{a}+
{i\over 2}{\stackrel{\leftarrow}{\partial}\over\partial{\tilde\phi}^{a}},
\phi^{a}-{i\over 2}{\stackrel{\leftarrow}{\partial}\over\partial
{\tilde\lambda}_{a}},{\bar c}_{a}+{\stackrel{\leftarrow}{\partial}\over
\partial {\tilde c}^{a}},c^{a}\bigr)
\vert_{{\tilde\lambda}=\lambda\atop
{{\tilde\phi}=\phi\atop {\tilde c}= c}}\cr}\eqn\viadana$$
The above equation is easily proven\refmark{10}~by Fourier-transforming
the functions ~$A$~and ~$B$. The Grassmannian variables ~$c^{a}$~ and
~${\bar c}^{a}$~ are the Wick symbols\refmark{10}~of fermionic creation and 
annihilation operators. This means that, for example,
the symbol~${\bar c}_{a}c^{b}=-c^{b}{\bar c}_{a}$~ represents the operator
~${\widehat{\bar c}}_{a}{\widehat c}^{b}$. The symbol for ~${\widehat c}^{b}
{\widehat{\bar c}}_{a}$~has an additional commutator term therefore.
In fact, eq.~\pordenone~yields
$$\eqalign{c^{a}\ast_{e}c^{b} \ = & \ c^{a}c^{b}\cr
{\bar c}_{a}\ast_{e}{\bar c}_{b}\ = & \ {\bar c}_{a}{\bar c}_{b}\cr
{\bar c}_{a}\ast_{e}c^{b}\ = & \ {\bar c}_{a}c^{b}\cr
c^{b}\ast_{e}{\bar c}_{a}\ = & \ c^{b}{\bar c}_{a}+
\delta^{b}_{a}\cr}\eqn\cogozzo$$
where it appears a normal-ordering term on the RHS of the last equation.
\chapter{DEFORMED SUPER-HAMILTONIAN: BOSONIC SECTOR}
Following the same strategy as in the classical case, we now try to find
a deformed super-Hamiltonian~${\widetilde{\cal H}}^{\hbar}\in Fun({\cal M}
_{8N})$~ which, with respect to the extended Moyal bracket, gives rise to the
same time-evolution of zero-forms as the standard Hamiltonian ~$H(\phi)$~with
respect to the standard Moyal bracket. Let us first try to solve the
deformed version of the zero-form equation
$$\bigl\{{\widetilde{\cal H}}^{\hbar},\varrho(\phi)\bigr\}_{emb}=\bigl\{
H(\phi),\varrho(\phi)\bigr\}_{mb}\eqn\casalmaggiore$$
For the time being we ignore the Grassmann variables. The p-form 
generalization of eq.~\casalmaggiore\break
will be investigated in section 6. Here
we look for a deformation of the bosonic part ~\paris\break only, 
~${\widetilde{\cal H}}_{B}=\lambda_{a}h^{a}(\phi)$. However, as we prove in
appendix A, the equation
$$\bigl\{\Haq_{B}(\lambda,\phi),\varrho(\phi)\bigr\}_{emb}=\bigl\{H(\phi),
\varrho(\phi)\bigr\}_{mb}\eqn\sangiovanni$$
does not possess any solution for
~$\Haq_{B}$. At this point classical mechanics does not offer any hint of
how to proceed and some new input is required. We shall weaken the condition
on~$\Haq_{B}$~by requiring that eq.~\sangiovanni~holds true only on the 
hypersurface where~$\lambda=0$. This is certainly a sensible choice, since
it is exactly the ~$\lambda=0$-hypersurface which is to be identified with the
standard phase-space ~${\cal M}_{2N}$, and only there the ordinary Moyal
formalism fixes the dynamics. So let us replace
eq.~\sangiovanni~ by
$${\cal P}\bigl\{\Haq_{B}(\lambda,\phi),\varrho(\phi)\bigr\}_{emb}=\bigl\{H(\phi),
\varrho(\phi)\bigr\}_{mb}\eqn\vho$$
where the projection operator ~${\cal P}$~acts on any function
~$F(\lambda,\phi)$~according to
$${\cal P}F(\lambda,\phi)=F(\lambda=0,\phi)\eqn\marcaria$$
Of course~$\lambda$~is set to zero in eq.~\vho~only after
the derivatives with respect to~$\lambda$~have been taken.
(In the language of Dirac's theory of constraints,
~$\lambda$~ is set to zero "weakly".)
In appendix A we show that the most general solution
to eq.~\vho~is given by
$$\Haq_{B}(\lambda,\phi)={1\over\hbar}\sinh\bigl[\hbar\lambda_{a}
\omega^{ab}\partial_{b}\bigr]H(\phi)+{1\over\hbar}
Y(\lambda,\phi)\eqn\castellucchio$$
Here ~$Y(\lambda,\phi)$~is an arbitrary function which is {\it even}
in ~$\lambda$~and of order~$\hbar^{2}$. Therefore, in the classical
limit, the second term on the RHS of ~\castellucchio~vanishes and the first one reproduces
the classical result:
$$ \lim_{\hbar\rightarrow 0}~\Haq_{B}(\lambda,\phi)=\lambda_{a}
\omega^{ab}\partial_{b}H=\lambda_{a}h^{a}={\widetilde{\cal H}}_{B}
(\lambda,\phi)\eqn\angeli$$
In our conventions ~$\Haq_{B}$~is explicitly dependent on the
deformation parameter~$\hbar$, but the extended Moyal bracket
~\pordenone~is not, so that formally, even in the classical limit,
all higher derivative terms are retained. Effectively these terms are all
irrelevant, however, since the classical~${\widetilde{\cal H}}_{B}$~is only
linear in ~$\lambda$.~(By rescaling ~$\lambda$~we could transfer the ~$\hbar$-
dependence from ~${\widetilde{\cal H}}$~to the bracket.)
\par
The arbitrary function ~$Y(\lambda,\phi)$~parametrizes the manner in which we
extrapolate the dynamics away from the ~$\lambda=0$-surface. It is instructive
to consider the following examples:
$$\eqalign{Y_{\pm}(\lambda,\phi) \ = & \ \pm\cosh\bigl[\hbar\lambda_{a}\omega
^{ab}\partial_{b}\bigr]~H(\phi)\cr
Y_{0}(\lambda,\phi)\ = & \ 0\cr}\eqn\villafranca$$
The power series of ~$Y_{0,\pm}$~start with a term~$\propto \hbar^{2}$~and
contain only even powers of~$\lambda$. It is remarkable that the resulting
super-Hamiltonians are given by the application of certain finite-difference
operators to the standard Hamiltonian ~$H(\phi)$. In fact, let us define, 
for any function ~$F(\phi)$, the following operations:
$$\eqalign{\bigl(D_{+}F\bigr)(\phi) \ = & \ {1\over\hbar}\bigl[F(\phi^{a}+
\hbar\omega^{ab}\lambda_{b})-F(\phi^{a})\bigr]\cr
\bigl(D_{-}F\bigr)(\phi) \ = & \ {1\over\hbar}\bigl[F(\phi^{a})-F(\phi^{a}-
\hbar\omega^{ab}\lambda_{b})\bigr]\cr
\bigl(D_{0}F\bigr)(\phi) \ = & \ {1\over 2\hbar}\bigr[F(\phi^{a}+\hbar
\omega^{ab}\lambda_{b})-F(\phi^{a}-\hbar\omega^{ab}\lambda_{b})
\bigr]\cr}\eqn\vicenza$$
Then eq.~\castellucchio~yields the following super-Hamiltonians
for the examples~\villafranca:
$$\eqalign{\Haq_{B~+}(\lambda,\phi) \ = & \ -D_{+}H(\phi)\cr
\Haq_{B~0}(\lambda,\phi) \ = & \ -D_{0}H(\phi)\cr
\Haq_{B~-}(\lambda,\phi) \ = & \ -D_{-}H(\phi)\cr}\eqn\mestre$$
The operators ~$D_{\pm,0}$~are strongly reminiscent of a forward,
backward and symmetric lattice derivative, respectively. They evaluate the
difference of some ~$F\in Fun({\cal M}_{2N})$~at two points with a finite 
separation ~$\hbar\omega^{ab}\lambda_{b}$. Here~$\lambda_{a}$~acts
as a parameter, external to ~${\cal M}_{2N}$, which defines how
the "links" of the lattice are imbedded into
~${\cal M}_{2N}$. In the classical limit the operators in ~\vicenza~
become identical and coincide with the directional derivative along
the vector ~$\omega^{ab}\lambda_{a}$:
$$\lim_{\hbar\rightarrow 0}D_{\pm,0}=\omega^{ab}\lambda_{b}
\partial_{a}\eqn\venezia$$
In order to further illuminate this lattice structure, let us look at the 
equations of motion of ~$\phi^{a}(t)$~and ~$\lambda_{a}(t)$.~From
$$\eqalign{{d\over dt}\phi^{a}(t) \ = & \ \bigl\{\phi^{a},\Haq_{B}\bigr\}_{emb}
={\partial\over\partial\lambda_{a}}{\Haq_{B}}(\lambda,\phi)\cr
{d\over dt}\lambda_{a}(t) \ = & \ \bigl\{\lambda_{a},\Haq_{B}\bigr\}_{emb}=
-{\partial\over\partial\phi^{a}}\Haq_{B}(\lambda,\phi)\cr}\eqn\livio$$
One easily obtains
$${d\over dt}\phi^{a}(t)=\cases{h^{a}(\phi^{b}+\hbar\omega^{bc}\lambda_{c})&,\cr
{1\over 2}\bigl[h^{a}(\phi^{b}+\hbar\omega^{bc}\lambda_{c})+h^{a}(\phi^{b}-
\hbar\omega^{bc}\lambda_{c})\bigr] &, \cr
h^{a}(\phi^{b}-\hbar\omega^{bc}\lambda_{c}) & ,\cr}\eqn\livia$$
where ~$\Haq_{B~+}$,~$\Haq_{B~0}$~and~$\Haq_{B~-}$~has been used, respectively.
Similarly
$${d\over dt}\lambda_{a}(t)=\cases{{1\over\hbar}\bigl[\partial_{a}H(\phi^{b}+
\hbar \omega^{bc}\lambda_{c})-\partial_{a}H(\phi^{b})\bigr] & ,\cr
{1\over 2\hbar}\bigl[\partial_{a}H(\phi^{b}+\hbar\omega^{bc}\lambda_{c})-
\partial_{a}H(\phi^{b}-\hbar\omega^{bc}\lambda_{c})\bigr] & ,\cr
{1\over\hbar}\bigl[\partial_{a}H(\phi^{b})-\partial_{a}H(\phi^{b}-\hbar
\omega^{bc}\lambda_{c})\bigr] & ,\cr}\eqn\fiorella$$
The trajectory~$\bigl(\phi^{a}(t),\lambda_{a}(t)\bigr)$~is the solution
of the coupled system of equations~\livia,~\fiorella. Contrary to
classical mechanics\refmark{11} where~$\lambda_{a}(t)$~does not influence 
the dynamics of ~$\phi^{a}(t)$, we see that here the ~$\phi$-dynamics depends
on ~$\lambda$. In a sense, ~$\phi^{a}(t)$~evolves on a "dynamical lattice"
on ~${\cal M}_{2N}$: by eq.~\fiorella~a fixed trajectory~$\phi^{a}(t)$~
leads to a certain solution~$\lambda_{a}(t)$~which defines a kind of
lattice which, in turn, governs the evolution of ~$\phi^{a}$~according to
~\livia. In fact, ~$\phi^{a}(t)$~does not feel the "drift force" ~$h^{a}$
at the point ~$\phi^{a}$~ as in the classical case, but rather at the "sites"
~$\phi^{b}\pm\hbar\omega^{bc}\lambda_{c}$~sitting at the ends of
the "link" given by the vector \break 
$\hbar\omega^{bc}\lambda_{c}$. Both in the classical limit ~$\hbar\rightarrow 0$~and upon
projection on the ~$\lambda=0$-surface (for ~$\hbar\ne 0$~) all
choices for ~$\Haq_{B}$~are equivalent, and eq.\livia~becomes
$${d\over dt}\phi^{a}(t)=h^{a}(\phi(t))\eqn\fiorello$$
Depending on whether we set ~$\hbar=0$~or ~$\lambda=0$~ at
~$\hbar\ne 0$~the interpretation of eq.~\fiorello~is different.
For ~$\hbar=0$, eq.~\fiorello~is Hamilton's classical equation
of motion. For ~$\lambda=0$,\break $\hbar\ne 0$~eq.\fiorello~is the equation for the
symbol~$\phi^{a}(t)$~of the Heisenberg operator~${\widehat\phi}^{a}(t)$.
This equation looks like Hamilton's equation because Heisenberg's equation has the same form as the classical one.
\par
Also the equations~\fiorella~become identical in the limit
~$\hbar\rightarrow 0$~where we recover the classical result
$${d\over dt}\lambda_{a}(t)=-\partial_{a}h^{b}(\phi(t))\lambda_{b}\eqn
\vittorio$$
It is important to note that for ~$\hbar\ne 0$~and for any choice of
~$\Haq_{B}$~eq.~\fiorella~always admits the solution~$\lambda_{a}(t)=0$
, for all ~$t$. This is essential for the consistency of our approach because 
it shows that the "constraint"~$\lambda_{a}=0$~is compatible with the
time evolution of the extended Moyal formalism: If the point
~$\bigl(\phi^{a},\lambda_{a}\bigr)(t)$~is on the hypersurface ($\lambda=0$) representing
~${\cal M}_{2N}$~at ~$t=0$, it will stay there at any time ~$t>0$.
\par
It would be interesting to keep~$\lambda\ne 0$~and from eq.~\livia~
and~\fiorella,~which are {\it modified Heisenberg equations of motion}
for the symbols~$\phi$~and $\lambda$, derive the associated
{\it modified Schr\"odinger equation}. The need for a modified Schr\"odinger
equation may arise at extremely high energies like in the realm of quantum gravity
where pure states might evolve into mixed ones. For sure this  modified
Schr\"odinger equation will bring new light on the field ~$\lambda$. 
This field is similar to the response field\Ref\alamos{P.C.Martin, E.D.Siggia 
and H.A.Rose, Phys.Rev.A 8 (1973) 423} of statistical mechanics and 
plays a central role in the fluctuation dissipation theorem (FDT). 
The fact that, to recover
standard quantum mechanics, we have to resctrict ourselves to ~$\lambda=0$  may imply that
the modified form of quantum mechanics violates the FDT. Also worth noticing is the fact
that the ~$\lambda$-field appeared already in stochastic processes
\Ref\revi{E.Gozzi, Suppl.Prog.Theor.Phys. 111 (1993) 115}~where it
effectively takes the place of the noise. So its presence now in modified
quantum Heisenberg evolution could indicate that this equation can be
turned into the standard Heisenberg equation coupled to noise on the line
of the recent interesting investigation reviewed in
\Ref\ghir{G.C.Ghirardi,~A.Rimini, in "{\it Sixty-Two Years of Uncertainty}",\nextline
A.I.Miller (Ed.), Plenum, New York, 1990}. We hope to come back to these topics
in the future.
\par
Let us now turn to less speculative topics and analyze in detail
the various choices of the ~$Y$-functions of ~\villafranca.
We shall show that the choice ~$Y=0$, leading to the symmetric
"lattice" derivative ~$D_{0}$, is singled out uniquely by a variety
of special features. First of all, looking at the second expression
contained in eqs.~\livia~and~\fiorella, we see that for that choice
 the equations of motion are invariant under the transformation
$$t\rightarrow -t~,~\phi^{a}\rightarrow \phi^{a}~,~\lambda_{a}\rightarrow
-\lambda_{a}~,~h^{a}\rightarrow -h^{a}\eqn\barile$$
For any other choice of ~$Y$~ this would not be the case. Second, let
us introduce the variables
$$\eqalign{X^{a}_{+} \ = & \ \phi^{a}+\hbar\omega^{ab}\lambda_{b}\cr
X^{a}_{-}\ = & \ \phi^{a}-\hbar\omega^{ab}\lambda_{b}\cr}\eqn\sgarbi$$
and let us derive their equations of motion. By adding and subtracting 
the second equation of ~\fiorella~and ~\livia~we find that
$$\eqalign{{d\over dt}X_{+}^{a}(t) \ = & \ h^{a}(X_{+}(t))\cr
{d\over dt}X^{a}_{-}(t) \ = & \ h^{a}(X_{-}(t))\cr}\eqn\ferrara$$
Remarkably,~$X^{a}_{+}$~and ~$X_{-}^{a}$~do not mix under time
evolution and, even for ~$\hbar\ne 0$~and ~$\lambda\ne 0$, they
separately obey the same equation as ~$\phi^{a}$~in the classical limit,
~see
eq.~\fiorello. This can be traced back to the fact that the symmetric 
Hamiltonian ~$\Haq_{B0}=-D_{0}H$~can be written as
$$\Haq_{B0}(\lambda,\phi)={1\over 2\hbar}\bigl[H(\phi^{a}-\hbar\omega^{ab}
\lambda_{b})-H(\phi^{a}+\hbar\omega^{ab}\lambda_{b})\bigr]\eqn\modugno$$
whence
$${\Haq}_{B0}={1\over 2\hbar}\bigl[H(X_{-}^{a})-H(X_{+}^{a})\bigr]\eqn\milva$$
and that the extended Moyal bracket decomposes similarly. For functions\break
~$A(X_{+},X_{-})$~and~$B(X_{+},X_{-})$~the star-product~\pordenone~
can be written as
$$A\ast_{e}B=A~exp\bigl[i\hbar\bigl({\stackrel{\leftarrow}{\partial}\over
\partial X^{a}_{-}}\omega^{ab}{\stackrel{\rightarrow}{\partial}\over
\partial X_{-}^{b}}-{\stackrel{\leftarrow}{\partial}\over
\partial X^{a}_{+}}\omega^{ab}{\stackrel{\rightarrow}{\partial}\over
\partial X_{+}^{b}}\bigr)\bigr]B\eqn\mina$$
Each one of the two terms in the bracket on the RHS of~\mina~has the
same structure as the operator~$\stackrel{\leftarrow}{\partial_{a}}
\omega^{ab}\stackrel{\rightarrow}{\partial_{b}}$~appearing in the ordinary
star product ~\vittos~with ~$\phi^{a}$~replaced by~$X^{a}_{-}$~and
~$X^{a}_{+}$, respectively. The overall factor is different, however.
This suggests to introduce the following modified star product
and Moyal bracket on ~$Fun({\cal M}_{2N})$:
$$\bigl(A\ast_{2}B\bigr)(\phi)=A(\phi)~exp\bigl[i\hbar~
\stackrel{\leftarrow}{\partial_{a}}\omega^{ab}\stackrel{\rightarrow}
{\partial_{b}}\bigr]~B(\phi)\eqn\reitano$$
$$\bigl\{A,B\bigr\}_{mb}^{(2)}={1\over 2i\hbar}\bigl(A\ast_{2}B-
B\ast_{2}A\bigr)\eqn\villa$$
They differ from the ~$\ast$~of ~\vittos~ and the ~$\bigl\{\cdot,
\cdot\bigr\}_{mb}$~of ~\vittu~only by the rescaling~$\hbar\rightarrow
2\hbar$.
\par
Let us now consider functions which depend only on either~$X_{-}^{a}$~
or~$X_{+}^{a}$. In a slight abuse of language we shall call them holomorphic and
antiholomorphic, respectively.\break These functions have the property that their
extended star-product can be expressed in terms of the modified star-product
~$\ast_{2}$~on ~$Fun({\cal M}_{2N})$. Eq.~\mina~implies that
$$\eqalign{A(X_{-})\ast_{e}~B(X_{-}) \ = & \ A(X_{-})~exp\bigl[+i\hbar 
{\stackrel{\leftarrow}{\partial}\over\partial X_{-}^{a}}\omega^{ab}
{\stackrel{\rightarrow}{\partial}\over\partial X_{-}^{b}}\bigr]B(X_{-})=A(\phi)
\ast_{2}B(\phi)\vert_{\phi=X_{-}}\cr
A(X_{+})\ast_{e}~B(X_{+})\ = & \ A(X_{+})~exp\bigl[-i\hbar {\stackrel
{\leftarrow}{\partial}\over\partial X_{+}^{a}}\omega^{ab}{\stackrel
{\rightarrow}{\partial}\over\partial X_{+}^{b}}\bigr]B(X_{+})=
B(\phi)\ast_{2}A(\phi)\vert_{\phi=X_{+}}\cr}\eqn\giocattolo$$
and that the ~$\ast_{e}$-product of a holomorphic with an antiholomorphic
function reduces to ordinary pointwise multiplication:
$$A(X_{\mp})\ast_{e}B(X_{\pm})=A(X_{\mp})B(X_{\pm})\eqn\bambola$$
This entails the following relations for the ~$em$-brackets
of (anti)-holomorphic functions
$$\eqalign{\bigl\{A(X_{\pm}),B(X_{\pm})\bigr\}_{emb} \ = & \ \mp2\hbar\bigl\{
A(\phi),B(\phi)\bigr\}^{(2)}_{mb}~\vert_{\phi=X_{\pm}}\cr
\bigl\{A(X_{\pm}),B(X_{\mp})\bigr\}_{emb} \ = & \ 0\cr}\eqn\carro$$
In particular it follows that
$$\eqalign{\bigl\{X_{-}^{a},X_{-}^{b}\bigr\}_{emb} \ = & \ +2\hbar\omega^{ab}\cr
\bigl\{X_{+}^{a},X^{b}_{+}\bigr\}_{emb} \ = & \ -2\hbar\omega^{ab}\cr
\bigl\{X_{+}^{a},X^{b}_{-}\bigr\}_{emb}\ = & \ 0\cr}\eqn\macchina$$
~\ie,~up to a factor of~$\pm 2\hbar$, the ~$emb$-algebra of ~$X_{-}^{a}$~
and ~$X_{+}^{a}$~coincides with the standard~$mb$-algebra~$\bigl\{\phi^{a},
\phi^{b}\bigr\}_{mb}=\omega^{ab}$. Thus the holomorphic and the antiholomorphic
functions form two closed and mutually commuting algebras with respect to the 
~$em$-bracket which are essentially equivalent to ~$Fun({\cal M}_{2N})$~
equipped with the standard Moyal bracket. This fact is important for
various reasons. First of all, it explains the simple form of the
equations~\ferrara~: neither the Hamiltonian~\milva~nor the ~$em$-brackets
couple the ~$X_{+}$-dynamics to the ~$X_{-}$-dynamics.
It has to be stressed, however, that this is true only for ~$Y=0$;~any
other choice would spoil the decoupling of ~$X_{+}$~and ~$X_{-}$.
\par
Another special property of the choice ~$Y=0$~is related to the
composition rule for two consecutive symplectic diffeomorphisms on
~${\cal M}_{2N}$. Denoting the generating functions for these transformations
~$G_{1}(\phi)$~and ~$G_{2}(\phi)$, the associated hamiltonian vector fields are
$h_{1,2}=(dG_{1,2})^{\sharp}\equiv\omega^{ab}\partial_{b}G_{1,2}\partial_{a}$~
and their Lie brackets form the following algebra~$$\bigl[h_{1},h_{2}\bigr]=
-h_{3}~~,~~h_{3}=\bigl(d\{G_{1},G_{2}\}_{pb}\bigr)^{\sharp}\eqn\gino$$
In the extended Poisson bracket formalism, this
equation is equivalent to\refmark{11}
$$\bigl\{{\widetilde{\cal H}}[G_{1}],{\widetilde{\cal H}}[G_{2}]\bigr\}_{epb}
={\widetilde{\cal H}}[\{G_{1},G_{2}\}_{pb}]\eqn\trattore$$
where ~${\widetilde{\cal H}}[G_{1,2}]$~is the classical super-Hamiltonian
~\pari~with ~$H(\phi)$~replaced by ~$G_{1,2}(\phi)$. It is important to
assess whether also the deformed (bosonic) super-Hamiltonians form a
closed algebra under the extended Moyal bracket. For~$Y=0$~we take
$$\Haq_{B0}[G_{1,2}]={1\over 2\hbar}\bigl[G_{1,2}(X_{-})-G_{1,2}(X_{+})
\bigr]\eqn\gru$$
so that eq.~\carro~implies
$$\eqalign{\bigl\{\Haq_{B0}[G_{1}],\Haq_{B0}[G_{2}]\bigr\}_{emb} \ = & \
\bigl({1\over 2\hbar}\bigr)^{2}\biggl(\bigl\{G_{1}(X_{-}),G_{2}(X_{-})
\bigr\}_{emb}+\bigl\{G_{1}(X_{+}),G_{2}(X_{+})\bigr\}_{emb}\biggr)\cr
\ = & \ \biggl(\bigl\{G_{1}(\phi),G_{2}(\phi)\bigr\}^{(2)}_{mb}\vert_{\phi=X_{-}}
- \bigl\{G_{1}(\phi),G_{2}(\phi)\bigr\}_{mb}^{(2)}\vert_{\phi=X_{+}}\biggr)\cr
\ = & \ {1\over 2 \hbar}\bigl[G_{3}(X_{-})-G_{3}(X_{+})\bigr]\cr}\eqn\organo$$
with ~$G_{3}=\bigl\{G_{1},G_{2}\bigr\}_{mb}^{(2)}$. We conclude that the
Hamiltonians~$\Haq_{B0}$~indeed form a closed algebra under the
~$em$-bracket:
$$\biggl\{\Haq_{B0}[G_{1}],\Haq_{B0}[G_{2}]\biggr\}_{emb}=\Haq_{B0}
\bigl[\{G_{1},G_{2}\}_{mb}^{(2)}\bigr]\eqn\ascensore$$
For~$Y\ne 0$~no comparable result can be proven in general. In the classical
limit the algebra~\ascensore~reduces to~\gino~, so we could call the
~\ascensore~ the algebra of the {\it quantum deformed} hamiltonian vector
fields.
\par
At first sight it might seem puzzling that the modified bracket
~$\bigl\{G_{1},G_{2}\bigr\}^{(2)}_{mb}$~appears on the RHS of~\ascensore.
~In fact, in the ordinary Moyal formalism, one would expect the standard
bracket~$\bigl\{G_{1},G_{2}\bigr\}_{mb}$~to appear. However, we have
to recall that ~$\bigl\{\Haq,{\cdot}\bigr\}_{emb}$~coincides with
~$\bigl\{H,\cdot\bigr\}_{mb}$~only after the projection on the
~$\lambda=0$-surface. Therefore iterated Moyal brackets
~$\bigl\{G_{1},\bigl\{G_{2},\cdot\bigr\}_{mb}\bigr\}_{mb}$~become
~${\cal P}\bigl\{\Haq_{B0}[G_{1}],{\cal P}\{\Haq_{B0}[G_{2}],\cdot\}_{emb}
\bigr\}_{emb}$~where the ~${\cal P}$~forbids a naive application of Jacobi's 
identity which would allow to combine the ~$\Haq$'s into one bracket.
Nevertheless, eq.~\ascensore~ shows that the ~$emb$-algebra of the
super-Hamiltonians closes even without intermediate projections. The only
change is the rescaling ~$\hbar\rightarrow 2\hbar$. Therefore,
once we have passed over from the ordinary Moyal formalism
to the extended one, if we interprete ~$2\hbar$~as the "physical" 
value of the deformation parameter, we may study the quantum analog of the
classical canonical transformations as ~$emb$-operations without
any subsequent projection.
\par
To close this section, we remark that the symmetric super-Hamiltonian
was also obtained by Marinov\Ref\rus{M.S.Marinov, Phys.Lett.A 153 (1991) 5}
in a completely different manner, namely by constructing a path-integral
solution to eq.~\tullio. An alternative derivation of Marinov's path-integral
is as follows. One starts from a density matrix operator~${\widehat\varrho}=
\ket{\psi}\bra{\psi}$~and expresses its time-evolution via the product of two Feynman path-integrals. Then,
after an appropriate change of variables, ~$symb({\widehat\varrho})$~can be 
seen to evolve according to Marinov's path-integral involving~$\Haq_{B0}(X_{+},
X_{-})$. In this derivation it becomes clear that the transformation
~\barile~which interchanges ~$X_{+}^{a}$~with ~$X_{-}^{a}$~corresponds
to interchanging vectors~$\ket{\psi}$~with dual vectors
~$\bra{\psi}$. Therefore eq.~\barile~ should be thought of as a kind of "modular
conjugation"~as it is explained in detail elsewhere\Ref\modu{E.Gozzi,
M.Reuter,~
"{\it Quantum deformed canonical transformations, \nextline 
$W_{\infty}$-algebras and
unitary transformations}",~UTS-DFT-93-11 preprint.}.
\chapter{DEFORMED SUPER-HAMILTONIAN: FERMIONIC SECTOR}
In the previous section we derived the bosonic Hamiltonian~$\Haq_{B}$~
which is the quantum deformed version of the classical super-Hamiltonian
~${\widetilde{\cal H}}_{B}=\lambda_{a}h^{a}$, or, equivalently of the 
hamiltonian vector field~$h^{a}\partial_{a}$. Thus, in a sense, ~$\Haq_{B0}$~
is the appropriate notion of a "quantum hamiltonian
vector field". In order to be able to deform more general geometric objects,
we have to generalize the fermionic part\foot{Also other authors
\Ref\junk{M.Blau,~E.Keski-Vakkuri,
A.Niemi, Phys.Lett.246B (1990) 92}
have  inserted Grassmannian variables in the standard
path-integral, but their techniques and goals were different from
the ones presented here.} of the super-Hamiltonian,
~${\widetilde{\cal H}}_{F}$. Then the complete
deformed hamiltonian
$$\Haq=\Haq_{B}+\Haq_{F}\in Fun({\cal M}_{8N})\eqn\nosari$$
will define a "{\it quantum Lie derivative} ~$L_{\hbar}$"~by an equation
analogous to ~\maiani:
$$\bigl\{\Haq,{\widehat\varrho}\bigr\}_{emb}=-(L_{\hbar}\varrho)^{\wedge}
\eqn\ferrero$$
Here the "hat-map~$(\cdot)^{\wedge}$"~is defined as in the classical case,
see eqs.~\piero~ and~\liana. Even though the natural multiplication
of the ~$c^{a}$'s~is via the extended star product~$\ast_{e}$~now, the classical
wedge product remains undeformed basically, because, by virtue of~\cogozzo,
$${\bar c}_{a}\ast_{e}{\bar c}_{b}\cdots\ast_{e}c^{e}\ast_{e} c^{f}\ast_{e}\cdots
={\bar c}_{a}{\bar c}_{b}\cdots c^{e}c^{f}\cdots\eqn\barilla$$
(A similar result was found in the third of ref.[6] in a different
framework.) 
\par
Because the new variables ~$c^{a}$~and~${\bar c}_{a}$~are not present in
conventional quantum mechanics, it is difficult to have any intuition on
how~$\Haq_{F}$~should be chosen. Of course we require that
$$\lim_{\hbar\rightarrow 0}\Haq_{F}={\widetilde{\cal H}}_{F}=i{\bar c}
_{a}\partial_{b}h^{a}c^{a}\eqn\nestle$$
but we are still left with a variety of possible choices
leading to different deformed calculi for ~$\hbar\ne 0$.
In this paper we study a very simple choice of~$\Haq_{F}$~which is 
inspired by our experience with the classical case. We require
~$\Haq_{F}$~to have the following two properties:
{\parindent=18pt
\item{\bf (F1)} The complete Hamiltonian~$\Haq=\Haq_{B}+\Haq_{F}$~is assumed to 
have the same BRS symmetry as its classical ancestor. This symmetry is generated
by the nilpotent charge~$Q=ic^{a}\lambda_{a}$. Both by ~$ep$-~and by ~$em$-
brackets it induces the following transformations on the fields 
($\epsilon$~is an anticommuting constant)}:
$$\eqalign{\delta\phi^{a} \ = & \ \epsilon c^{a}\cr
\delta{\bar c}_{a} \ = & \ i\epsilon \lambda_{a}\cr
\delta c^{a} \ = & \ 0 = \delta\lambda_{a}\cr}\eqn\bologna$$
{\parindent=18pt
\item{\bf (F2)} We assume that ~$\Haq_{F}$~is bilinear in ~$c^{a}$~
and ~${\bar c}_{a}$,~\ie,~that, for some function~$W^{a}_{b}$,}
$$\Haq_{F}=i{\bar c}_{a}~W^{a}_{b}(\lambda,\phi)~c^{b}\eqn\nocera$$
We show in appendix B that for any bosonic Hamiltonian of the form
$$\Haq_{B}(\lambda,\phi)={\cal F}(\lambda_{a}\omega^{ab}\partial_{b})
H(\phi)\eqn\battipaglia$$
(with ~${\cal F}(x)=x$~in the classical limit) the two conditions 
{\bf (F1)} and {\bf (F2)} fix the
fermionic piece ~$\Haq_{F}$~
uniquely. The answer we find is
$${\widetilde{\cal H}^{\hbar}}={\cal F}(\lambda\omega\partial)H(\phi)+i
{\bar c}_{a}\omega^{ac}\partial_{b}\partial_{c}{{\cal F}
(\lambda\omega\partial)\over (\lambda\omega\partial)}
H(\phi)c^{b}\eqn\ruth$$
where ~$\lambda\omega\partial\equiv\lambda_{a}\omega^{ab}\partial_{b}$. In
section 5 we argued that a preferred choice for ~$\Haq_{B}$~is the 
symmetric one,~\ie,~$Y=0$~in eq.~\castellucchio. Therefore we shall use from now
on:
$${\cal F}(x)={1\over\hbar}\sinh(\hbar x)\eqn\impaziente$$
Note that ~${\cal F}(x)=x+O(\hbar^{2})$.
\par
The Hamiltonian ~\ruth~is BRS invariant by construction. It is remarkable
that it is also invariant under the transformations generated by all the other
charges listed in~\valentina~:
$$\bigl\{\Haq,\Omega\bigr\}_{emb}=0\eqn\veloce$$
$$\Omega\equiv Q,{\bar Q},Q_{g},K,{\bar K}\eqn\lento$$
It is particularly interesting that the charge ~$K$~is conserved. In fact,
in the classical case, it was shown\refmark{11}that this conservation is 
equivalent to Liouville's theorem. In quantum mechanics we have an analogous "formal" conservation because the
Heisenberg equations of motion (or the corresponding equations for the
symbols) have the same form as the classical Hamiltonian equations. Of course
this does not mean that in quantum mechanics we have a Liouville theorem 
(\ie,~that the volume of phase-space is conserved under time evolution). 
The formal analogy between Heisenberg and Hamilton equations holds
quantum mechanically only at the operatorial level  and not at the level of averaged
quantities. 
\par
The generators ~$\Omega$~in equation~\lento~ act on ~$A\in Fun({\cal M}_{8N})$~via the
~$em$-brackets:
$$\delta A=i\bigl\{\epsilon \Omega,A\bigr\}_{emb}\eqn\bello$$
As the ~$\Omega$'s are only quadratic in ~$\phi^{a},\lambda_{a},\cdots$,
the expressions for ~$\delta\phi^{a}$,~$\delta\lambda_{a}$,$\cdots$
are the same as in the classical case\refmark{11}. The ~$\Omega$'s obey
the following algebra:
$$\bigl\{Q, Q\bigr\}_{emb}=\bigl\{Q,{\bar Q}\bigr\}_{emb}=\bigl\{{\bar Q},
{\bar Q}\bigr\}_{emb}=0\eqn\barbiere$$
$$\eqalign{ \  i & \ \bigl\{Q_{g},Q\bigr\}_{emb}=Q~~~~~,~~~i\bigl\{Q_{g},
{\bar Q}\bigr\}_{emb}=-{\bar Q}\cr
\  i & \ \bigl\{ K,Q\bigr\}_{emb}=0~~~~~~~,~~~i\bigl\{K,{\bar Q}\bigr\}_{emb}
={\bar Q}\cr
\ i & \ \bigl\{{\bar K}, Q\bigr\}_{emb}={\bar Q}~~~~~~,~~~i\bigl\{{\bar
K},{\bar Q}\bigr\}_{emb}=0\cr
\ i & \ \bigl\{Q_{g}, K\bigr\}_{emb}=2K~~~,~~~i\bigl\{Q_{g},{\bar K}\bigr\}_{emb}=
-2{\bar K}\cr
\ i & \ \bigl\{K,{\bar K}\bigr\}_{emb}=Q_{g}+N\cr}\eqn\barba$$
This is the same ~$ISp(2)$~algebra as in the classical case\refmark{11}
except for the last equation which was ~$i\bigl\{K,{\bar K}\bigr\}_{epb}
=Q_{g}$~there. The new term~$+N$~is easily removed reordering the
~$c$'s in the "ghost-charge" ~$Q_{g}$. We replace ~$Q_{g}=c^{a}
{\bar c}_{a}\equiv c^{a}\ast_{e}{\bar c}_{a}-2N$~by the symmetric combination
$${\widetilde Q}_{g}={1\over 2}\bigl(c^{a}\ast_{e}{\bar c}_{a}-{\bar c}_{a}
\ast_{e} c^{a}\bigr)=Q_{g}+N\eqn\mento$$
so that the ~$+N$~disappears from the last equation without changing the
other ones.
\par
Comparing the Hamiltonian ~\ruth~ to its classical limit given in eqs.\paris,
~\parisi, we observe that the only effect of the quantum deformation consists of
replacing the classical Hamiltonian ~$H(\phi)$~by its "{\it quantum lift}"
$$H_{\hbar}(\lambda,\phi)={{\cal F}(\lambda\omega\partial)\over (\lambda\omega
\partial)}H(\phi)\eqn\menta$$
where ~${\cal F}$~is given by~\impaziente.
In fact, we can re-write~\ruth~as
$$\Haq=\lambda_{a}h^{a}_{\hbar}(\lambda,\phi)+i{\bar c}_{a}~
\partial_{b}h^{a}_{\hbar}
(\lambda,\phi)~c^{b}\eqn\rabarbaro$$
with the following deformed components of the hamiltonian vector field
$$h^{a}_{\hbar}(\lambda,\phi)=\omega^{ab}\partial_{b}H_{\hbar}(\lambda,
\phi)\eqn\zucca$$
We refer to ~$H_{\hbar}(\lambda,\phi)$~as the quantum {\it lift} of ~$H(\phi)$~
because, contrary to ~$H$~, it depends also on ~$\lambda$
so it is {\it lifted} from the ~$Fun({\cal M}_{2N})$~to
~$Fun({\cal M}_{4N})$. In our interpretation the bosonic subspace
~${\cal M}_{4N}\equiv\bigl\{(\lambda,\phi)\bigr\}$~of ~${\cal M}_{8N}$~is 
identified with the tangent bundle ~$T{\cal M}_{2N}$~over standard phase-space. Thus
the deformation lifts the Hamiltonian from a function on ~${\cal M}_{2N}$~to
a function on ~$T{\cal M}_{2N}$. Similarly,~$h^{a}\partial_{a}$~is lifted to 
a vector field~$h^{a}_{\hbar}\partial_{a}$~on ~${\cal M}_{4N}$. This vector
field is horizontal in the sense that a generic vector field on~${\cal M}_{4N}$~has the structure
$$v=v_{1}^{a}{\partial\over\partial\phi^{a}}+v_{2}^{a}{\partial\over\partial
\lambda_{a}}\eqn\cocomero$$
but the ~${\partial\over\partial\lambda_{a}}$-piece is missing in
~$h^{a}_{\hbar}\partial_{a}$. 
\par
Eq.~\menta~can be rewritten in a rather 
intriguing manner. Exploiting the identity
$${\sinh (\hbar x)\over (\hbar x)}={1\over 2}\int_{-1}^{1}~ds~
exp[-\hbar x s]\eqn\rosita$$
and the fact that ~$exp[-\hbar(\lambda\omega\partial)s]$~is a shift 
operator on ~${\cal M}_{2N}$, we arrive at
$$H_{\hbar}(\lambda,\phi)={1\over 2}~\int_{-1}^{+1}~ds~H(\phi^{a}+\hbar
\omega^{ab}\lambda_{b}s)\eqn\rosa$$
We see that the deformed Hamiltonian ~$H_{\hbar}$~is a kind
of "average" of ~$H$, which is performed along a straight line
centered at ~$\phi^{a}$~and~connecting ~$(\phi^{a}-\hbar
\omega^{ab}\lambda_{b})$~to~$(\phi^{a}+\hbar
\omega^{ab}\lambda_{b})$,~\ie,~along a "link" of the lattice mentioned
earlier. As the length of the link is proportional to ~$\hbar$, the
semiclassical limit has an obvious geometrical interpretation in this language:
for ~$\hbar\rightarrow 0$~the averaging is over very short line segments 
so that ~$H_{\hbar}(\lambda,\phi)\approx~H(\phi)$, and the dependence on
~$\lambda$~disappears. Conversely, in this framework it is easy to understand
where the non-local features of quantum mechanics come from: as we turn
on ~$\hbar$, the relevant Hamiltonian is not ~$H(\phi)$~anymore, but
~$H_{\hbar}(\lambda,\phi)$~which is a "smeared" version of ~$H(\phi)$. It
is remarkable that this smearing is performed along a one-dimensional
line only, and that the orientation of this line is determined by 
$\lambda_{a}$~ which, as we mentioned before,
plays a role similar to the response field\refmark{20} in statistical mechanics.
\par
We remark that also the deformed super-Hamiltonian is a pure BRS variation.
It is in fact easy to show that
$$\Haq=\bigl\{Q,h^{a}_{\hbar}{\bar c}_{a}\bigr\}_{emb}\equiv\bigl\{Q,
{\widehat h}_{\hbar}\bigr\}_{emb}\eqn\rose$$
with ~$h^{a}_{\hbar}$~given by~\zucca. This suggests that one
might obtain a one-dimensional topological field theory once
~$\Haq$~is inserted into a path-integral with BRS invariant boundary
conditions on the line of what we\refmark{18} did in the classical case.
\par
Having added a Grassmannian piece to ~$\Haq_{B}$~we have to make sure that the
flow induced by ~$\Haq$~leaves invariant the ~$\lambda=0$-hypersurface
which we identified with the standard phase-space ~${\cal M}_{2n}$.
The equations of motion,~${d\over dt}A=\{A,\Haq\}_{emb}$, for ~$A=\phi^{a},
\lambda_{a},c^{a},{\bar c}_{a}$,~respectively, read
$$\eqalign{{\dot\phi}^{a}={1\over 2}\bigl[h^{a}(\phi+\hbar\omega\lambda)\ + & \
h^{a}(\phi-\hbar\omega\lambda)\bigr]\cr
\ + & \ {i\over 2}\hbar\omega^{ab}\omega^{cd}\int_{-1}^{+1}~ds~s~\partial_{b}
\partial_{c}\partial_{e}H(\phi+\hbar\omega\lambda s)~{\bar c}_{d}
c^{e}\cr}\eqn\margherita$$
$$\eqalign{{\dot\lambda}_{a}={1\over 2\hbar}\bigl[\partial_{a}H(\phi+
\hbar\omega\lambda) \ - & \ \partial_{a}H(\phi-\hbar\omega\lambda)\bigr]\cr
\ - & \ {i\over 2}~\int_{-1}^{+1}~ds~\partial_{a}\partial_{b}~h^{c}(\phi+
\hbar\omega\lambda s)~{\bar c}_{c}~c^{b}\cr}\eqn\viola$$
$${\dot c}^{a}={1\over 2}~\int_{-1}^{+1}~ds~\partial_{b}h^{a}
(\phi+\hbar\omega\lambda s)c^{b}\eqn\geranio$$
$${\dot{\bar c}}_{a}=-{1\over 2}~\int_{-1}^{+1}~ds~\partial_{a}h^{b}
(\phi+\hbar\omega\lambda s)~{\bar c}_{b}\eqn\oleandro$$
$\lambda_{a}\equiv 0$~is not in general a solution of the coupled set
of equations because of the Grassmannian piece on the RHS of ~\viola;
the bilinear ~${\bar c}c$~acts as a source for ~$\lambda$. In order to
eliminate this term, we supplement the "constraint"~$\lambda=0$~
by its Grassmannian  counterpart ~${\bar c}_{a}=0$. It is easy to see that ~$\lambda_{a}(t)=0$~
and ~${\bar c}_{a}(t)=0$~is always a solution of the above equations
of motion, and that in this case the equation for ~$\phi^{a}$~
and ~$c^{a}$~reduce to Hamilton's and Jacobi's equation,
respectively:
$$\eqalign{{\dot\phi}^{a} \ = & \ h^{a}(\phi)\cr
{\dot c}^{a} \ = & \ \partial_{b}h^{a}(\phi)~c^{b}\cr}\eqn\tiglio$$
We conclude that the hypersurface consisting of the points
~$\bigl(\phi,\lambda=0,c,{\bar c}=0\bigr)\in {\cal M}_{8N}$~is
preserved under the hamiltonian flow, and that the symbol ~$\phi^{a}(t)$~
correctly follows the Heisenberg dynamics of ~${\widehat\phi}^{a}(t)$.
The new feature is the symbol ~$c^{a}(t)$~parametrizing nearby
~$\phi^{a}$-trajectories. Stated differently, the function ~$c^{a}(t)$~
is an element of the tangent space to the space of bosonic paths
$\phi^{a}(t)$\refmark{11}.
\chapter{THE DEFORMED EXTERIOR CALCULUS}
In this section we apply the extended Moyal deformation outlined in
sections 4,~5 and 6 to the hamiltonian formulation of the exterior calculus
which we introduced in section 3. Even in the deformed case, the charges
~$\Omega\equiv Q, {\bar Q}, Q_{g},K,{\bar K}$~have vanishing ~$em$-brackets
with~$\Haq$~and form a closed ~$ISp(2)$~algebra. This enables us to set
up a deformed exterior calculus by paralleling the classical construction.
\par
The "hat map" is defined as in the classical case. It replaces ~$\partial_{a}$~
and ~$d\phi^{a}$~by~${\bar c}_{a}$~and ~$c^{a}$, and it maps tensors of the type
~\piero~to functions ~${\widehat T}\in Fun({\cal M}_{8N})$~as given in ~\liana.
It will be helpful to rewrite eq.~\liana~ as
$${\widehat T}=T^{b_{1}\cdots b_{q}}_{a_{1}\cdots a_{p}}(\phi)~{\bar c}_{b_{1}}
\ast_{e}~{\bar c}_{b_{2}}\cdots\ast_{e}{\bar c}_{b_{q}}\ast_{e}c^{a_{1}}
\ast_{e}\cdots\ast_{e}c^{a_{p}}\eqn\lianas$$
where ~\barilla~ has been used. The algebra of functions~$Fun({\cal M}_{8N})$~
is equipped with the star product ~$\ast_{e}$~or, equivalently, the extended
Moyal bracket. The operations of the deformed tensor calculus will be implemented by
~$em$-brackets between elements of ~$Fun({\cal M}_{8N})$~representing
tensors on ~${\cal M}_{8N}$~and the ~$ISp(2)$~generators. In order to
describe the deformed calculus, we consider the special tensors
~$v,\alpha,F^{(p)}$~and~$V^{(p)}$~defined in eq.~\maria. We {\it define}
a quantum exterior derivative of p-forms~$F^{(p)}$~as the extended Moyal
bracket with the BRS charge ~$Q$
$$\bigl(dF^{(p)}\bigr)^{\wedge}=i\bigl\{Q,
{\widehat F}^{(p)}\bigr\}_{emb}\eqn\arturo$$
In particular, for zero forms
$$\bigl(df\bigr)^{\wedge}=i\bigl\{Q,f(\phi)\bigr\}_{emb}=\partial_{a}f(\phi)
c^{a}\eqn\trento$$
Because ~$Q=ic^{a}\lambda_{a}\equiv i c^{a}\ast_{e}\lambda_{a}$~is quadratic
there is no difference between ~$\{Q,\cdot\}_{emb}$~and the classical
~$\{Q,\cdot\}_{epb}$. However, we can extend the definition of~"$d$"~to any 
~${\widehat T}\in Fun({\cal M}_{8N})$~of the form~\lianas~with the 
coefficients possibly depending also on ~$\lambda$:
$$d{\widehat T}=i\bigl\{Q,{\widehat T}\bigr\}_{emb}\eqn\bolzano$$
Even through this formula looks classical, there is an important difference
with respect to the Leibniz rule obeyed by the deformed~"$d$": it is a 
graded derivation on the algebra~$\bigl(Fun({\cal M}_{8N}),\ast_{e}\bigr)$
~instead of the classical ~$\bigl(Fun({\cal M}_{8N}),\cdot\bigr)$. Eq.
~\mantova~implies that
$$\bigl\{Q,{\widehat T}_{1}\ast_{e}{\widehat T}_{2}\bigr\}_{emb}=\bigl\{Q,
{\widehat T}_{1}\bigr\}_{emb}\ast_{e}{\widehat T}_{2}+(-1)^{[T_{1}]}{\widehat
T}_{1}\ast_{e}\bigl\{Q,{\widehat T}_{2}\bigr\}_{emb}\eqn\adige$$
There is no simple way of expressing the exterior derivative of the pointwise
product ~${\widehat T}_{1}\cdot{\widehat T}_{2}$, which, in this context,
is a quite unnatural object.
It is easy to see that the relation\break $\{Q,Q\}=0$~implies the nilpotency of
~$d$,~\ie,~$d^{2}=0$.
\par
Similar remarks apply to the exterior coderivative
$$\bigl({\bar d}V^{(p)}\bigr)^{\wedge}=i\bigl\{{\bar Q},
{\widehat V}^{(p)}\bigr\}_{emb}\eqn\olio$$
On zero-forms it acts as ~${\bar d}f=(df)^{\sharp}=\omega^{ab}\partial_{b}
f\partial_{a}$,~like in the classical case. Note that, as a consequence of
~$\{Q,{\bar Q}\}_{emb}=0$, we have ~$d{\bar d}+{\bar d}d=0$. (This is different
from Riemannian geometry, where the corresponding anticommutator yields the 
Laplace-Beltrami operator.)
\par
Further graded derivations obeying a Leibniz rule similar to~\adige~
include the interior products of p-forms and p-vectors with vectors
and 1-forms, respectively:
$$\eqalign{\bigl(i(v)F^{(p)}\bigr)^{\wedge} \ = & \ i\bigl\{{\widehat
v},{\widehat F}^{(p)}\bigr\}_{emb}\cr
\bigl(i(\alpha)V^{(p)}\bigr)^{\wedge}\ = & \ i\bigl\{{\widehat\alpha},
{\widehat V}^{(p)}\bigr\}_{emb}\cr}\eqn\mangano$$
They,~too, can be extended to any ~${\widehat T}\in Fun({\cal M}_{8N})$.
The maps relating vectors ~$v$~to 1-forms~$v^{\flat}$~and 1-forms~$\alpha$
to vectors ~$\alpha^{\sharp}$~are realized as the brackets with
~$K$~and ~${\bar K}$~again:
$$\eqalign{\bigl(v^{\flat}\bigr)^{\wedge} \ = & \ i\bigl\{K,{\widehat v}\bigr
\}_{emb}\cr
\bigl(\alpha^{\sharp}\bigr)^{\wedge}\ = & \ i\bigl\{{\bar K},{\widehat\alpha}
\bigr\}_{emb}\cr}\eqn\porrati$$
Finally we turn to the quantum Lie derivative ~$L_{h}$~along the hamiltonian
vector field. Choosing a fermionic Hamiltonian~$\Haq_{F}$~amounts to deciding
for a specific form of the Lie derivative. For any
~${\widehat T}\in Fun({\cal M}_{8N})$~we define
$$L_{h}{\widehat T}=-\bigl\{\Haq,{\widehat T}\bigr\}_{emb}\eqn\rifiuti$$
If ~${\widehat T}$~represents a tensor of the form~\lianas~its Lie derivative, considered as an operation acting on tensors
on ~${\cal M}_{2N}$, is obtained by projecting ~\rifiuti~
on the ~$\lambda=0$-surface
$$\bigl(L_{h}T\bigr)^{\wedge}=-{\cal P}\bigl\{\Haq,{\widehat T}\bigr\}_{emb}
\eqn\verde$$
Clearly~$L_{h}$~of ~\rifiuti~obeys
$$L_{h}\bigl({\widehat T}_{1}\ast_{e}{\widehat T}_{2}\bigr)=\bigl(L_{h}
{\widehat T}_{1}\bigr)\ast_{e}{\widehat T}_{2}+{\widehat T}_{1}\ast_{e}
\bigl(L_{h}{\widehat T}_{2}\bigr)$$
but the projected quantity~\verde~has no simple composition properties
anymore. We can show that~\rifiuti,~without projection to ~$\lambda=0$,
obeys
$$L_{h}=di(h_{\hbar})+i(h_{\hbar})d\eqn\rosso$$
which implies that ~$dL_{h}=L_{h}d$.
The proof of ~\rosso~ makes use of ~\arturo~,~\mangano,~\rose~ and the Jacobi
identity for the ~$em$-bracket:
$$\eqalign{di(h_{\hbar}){\widehat T}+i(h_{\hbar})d{\widehat T}\ = & \  i
\bigl\{Q,i\{{\widehat h}_{\hbar},{\widehat T}\}_{emb}\bigr\}_{emb}\cr
\ ~~~+ & \ i\bigl\{{\widehat h}_{\hbar},i\{Q,{\widehat T}\}_{emb}\bigr\}_{emb}\cr
\ = & \ -\bigl\{\{Q,{\widehat h}_{\hbar}\}_{emb},{\widehat T}\bigr\}_{emb}\cr
\ = & \ -\bigl\{\Haq,{\widehat T}\bigl\}_{emb}=L_{h}
{\widehat T}\cr}\eqn\giallo$$ 
Eq.~\rosso~ suggests the introduction of the operation
$$I(v)\equiv i(v_{\hbar})\eqn\viola$$
consisting of the lift~$v\mapsto v_{\hbar}$~followed by the 
contraction ~"$i$". Here ~$v=v^{a}(\phi)\partial_{a}$~is any (not necessarily
hamiltonian) vector field and
$$v_{\hbar}(\lambda,\phi)={{\cal F}(\lambda\omega\partial)\over(\lambda
\omega\partial)}v^{a}(\phi)\eqn\arancione$$
is its quantum lift. Then, for arbitrary vector fields,
$$L_{v}=dI(v)+I(v)d\eqn\arancio$$
with ~$L_{v}$~defined by ~\rifiuti~ where ~$\Haq$~is given by
~\rabarbaro~with ~$h^{a}_{\hbar}$~replaced by ~$v^{a}_{\hbar}$. In terms of
~$em$-brackets we have
$$I(v){\widehat T}=i\biggl\{{{\cal F}(\lambda\omega\partial)\over (\lambda\omega
\partial)}{\widehat v},{\widehat T}\biggl\}_{emb}\eqn\pere$$
As an example we present the contraction of ~${\widehat v}$~with ~${\widehat F}
^{(p)}$~as defined in eq.~\maria~. Using eq.~\viadana~it is easy to show 
that for any ~$\lambda$
$$I_{v}{\widehat F}^{(p)}={1\over (p-1)!}{\cal G}\bigl([\lambda_{a}-{1\over 2}
\stackrel{1}{\partial_{a}}]\omega^{ab}\stackrel{2}{\partial_{b}}\bigr)~v^{a}
(\phi_{1})~F^{(p)}_{a~a_{2}\cdots a_{p}}(\phi_{2})\vert_{\phi_{1,2}=\phi}
~c^{a_{2}}\cdots c^{a_{p}}\eqn\mele$$
where ~$${\cal G}(x)\equiv{{\cal F}(x)\over x}\equiv{\sinh(\hbar x)\over 
(\hbar x)}\eqn\dizionario$$
approaches unity in the classical limit so that~$I_{v}\rightarrow i_{v}$.
For the special case of a hamiltonian vector field ~$v^{a}=\omega^{ab}
\partial_{b}G\equiv(dG)^{\sharp a}$~contracted with a gradient~$F^{(1)}_{a}=
\partial_{a}f=(df)_{a}$,~the~$\lambda=0$-projection of ~\mele~ can be expressed 
as a standard Moyal bracket:
$${\cal P}~I\bigl((dG)^{\sharp}\bigr)~{\widehat {df}}=
\bigl\{f,G\bigr\}_{mb}\eqn\francia$$
In this case the ~$\bigl(\hbar x\bigr)^{-1}$-piece from ~${\cal G}$~
cancels the derivatives in ~$v^{a}$~and ~$F^{(1)}_{a}$~and we are left
with~$\sin\bigl({\hbar\over 2}\stackrel{1}{\partial_{a}}\omega^{ab}\stackrel
{2}{\partial_{b}}\bigr)$~leading to a normal Moyal bracket. Hence Moyal's 
equation of motion for the time-evolution of pseudodensities may be re-written
as
$$-\partial_{t}\varrho=\bigl\{\varrho,H\bigr\}_{mb}={\cal P}~I(h)~{\widehat{
d\varrho}}\eqn\inghilterra$$
which generalizes the classical result\refmark{11,14}
$$-\partial_{t}\varrho = \bigl\{\varrho,H\bigr\}_{pb}=i(h)~{\widehat {d\varrho}}
\eqn\lichtenstein$$
We stress once more that the RHS of eq.~\lichtenstein~is automatically 
independent of ~$\lambda$~ but not so the RHS of eq.\inghilterra. Using eq.
~\viadana~we can convince ourselves that eq.~\inghilterra~is equivalent to
$$\bigl\{\varrho,H\bigr\}_{mb}={\cal P}~h^{a}_{\hbar}(\lambda,\phi)\ast_{e}
~\partial_{a}\varrho(\phi)\eqn\austria$$
Comparing this to the classical formula~$\{\varrho,H\}_{pb}
=h^{a}\partial_{a}\varrho$~one might be tempted to consider the 
pseudodifferential operator
$${\cal D}(h)\equiv h^{a}_{\hbar}(\lambda,\phi)~
\partial_{a}\ast_{e}\eqn\altensteig$$
acting on ~$Fun({\cal M}_{4N})$, as the deformed version of the first order
operator ~$h^{a}\partial_{a}$~to which ~${\cal D}(h)$~reduces in the classical
limit. However, this interpretation is unnatural for the following reason.
The classical vector fields~$h^{a}\partial_{a}$~form a closed algebra~\gino,
but the algebra of the ~${\cal D}$'s does not close. For two hamiltonian
vector fields ~$h_{1,2}^{a}=\omega^{ab}\partial_{b}G_{1,2}$~one obtains
$$\bigl[{\cal D}(h_{1}),{\cal D}(h_{2})\bigr]={\cal D}(h_{3})+i\bigl\{
h^{a}_{1,\hbar},h^{b}_{2,\hbar}\bigr\}_{emb}\ast_{e}\partial_{a}\partial_{b}
\eqn\tubingen$$
with
$$h^{a}_{3,\hbar}\equiv h^{b}_{1,\hbar}\ast_{e}\partial_{b}h^{a}_{2,\hbar}
-h^{b}_{2,\hbar}\ast_{e}\partial_{b}h^{a}_{1,\hbar}\eqn\insalata$$
The first term on the RHS of ~\tubingen~resembles the familiar Lie bracket, 
but the second one is new and spoils the closure of the algebra. It vanishes in
the limit~$\hbar\rightarrow 0$. When we try to re-express the ~$em$-bracket
of the associated Hamiltonian ~$\Haq_{B0}[G_{1,2}]=\lambda_{a}
h^{a}_{1,2;\hbar}$~in terms of ~$h^{a}_{1,2;\hbar}$~we obtain a formula similar
to ~\tubingen~suffering from the same problem:
$$\bigl\{\Haq_{B0}[G_{1}],\Haq_{B0}[G_{2}]\bigr\}_{emb}=-h^{a}_{3,\hbar}
\lambda_{a}+\bigl\{h^{a}_{1,\hbar},h^{b}_{2,\hbar}\bigr\}\ast_{e}\lambda_{a}
\lambda_{b}\eqn\mostarda$$
In Section 5 we have seen that the algebra of the ~${\Haq}_{B}$'s does
indeed close. Eq.~\ascensore~can be re-written as
$$\bigl\{h^{a}_{1,\hbar}\lambda_{a},h^{b}_{2,\hbar}\lambda_{b}\bigr\}_{emb}=v_{12}^{a}
\lambda_{a}\eqn\finocchi$$
where the new vector field ~$v^{a}_{12}$~is the quantum lift
of~$\omega^{ab}\partial_{b}\bigl\{G_{1},G_{2}\bigr\}^{(2)}_{mb}$:
$$v_{12}^{a}={{\cal F}(\lambda\omega\partial)\over (\lambda\omega\partial)}
~\bigl[\bigl\{h^{a}_{1},G_{2}\bigr\}^{(2)}_{mb}+\bigl\{G_{1},h^{a}_{2}\bigr\}
^{(2)}_{mb}\bigr]\eqn\zucchini$$
Composing two consecutive canonical transformations by the rule~\zucchini~
leads to a closed algebra, but using eq.~\insalata~it does not. The lesson 
to be learned from this is that, in the deformed case, it is unnatural to
decompose the hamiltonian vector field as the product of the "components"
~$h^{a}_{\hbar}$~with "basis elements" ~$\partial_{a}$~or ~$\lambda_{a}$.
As ~$h^{a}_{\hbar}(\lambda,\phi)$~contains arbitrary powers of ~$\lambda$~,
it is meaningless to separate off a single factor of ~$\lambda$. (Clearly
the situation is different in the classical case where ~${\widetilde{\cal H}}
_{B}=h^{a}\lambda_{a}$~is always linear in ~$\lambda$.) In quantum mechanics it
seems appropriate to call ~$\Haq_{B}$~{\it as a whole}~the "hamiltonian
vector field". Generally speaking it is a complicated function on the tangent
bundle over ~${\cal M}_{2N}$; only in the classical limit it becomes a vector
field on ~${\cal M}_{2N}$. In more physical terms, the higher powers of ~
$\lambda_{a}$~or ~$\partial_{a}$~are an expression of the nonlocal nature of
quantum mechanics.
\par
As an example we give the explicit form of the quantum Lie derivative
for tensors of the form~\lianas. The ~$em$-bracket of ~${\widehat T}$~with the
super-Hamiltonian is given by
$$\eqalign{\bigl\{\Haq,{\widehat T}\bigr\}_{emb} \ = & \ \biggl(i\biggl[{\cal F}
([\lambda+{i\over 2}\stackrel{1}{\partial}]\omega\stackrel{2}{\partial})-
{\cal F}([\lambda-{i\over 2}\stackrel{1}{\partial}]\omega\stackrel{2}
{\partial})\biggr]~T^{b_{1}\cdots b_{q}}_{a_{1}\cdots a_{p}}(\phi_{1})
~H(\phi_{2})\cr
\ + & \ \sum_{j=1}^{q}{\cal G}(\lambda\omega\stackrel{2}{\partial}
+{i\over 2}\stackrel{1}{\partial}\omega\stackrel{2}{\partial})~T^{b_{1}\cdots
b_{j-1}eb_{j+1}\cdots b_{q}}_{a_{1}\cdots a_{p}}(\phi_{1})
~\stackrel{2}{\partial_{e}}h^{b_{j}}(\phi_{2})\cr
\ - & \ \sum_{j=1}^{p}{\cal G}(\lambda\omega\stackrel{2}{\partial}+
{i\over 2}\stackrel{1}{\partial}\omega\stackrel{2}{\partial})~T^{b_{1}\cdots
b_{q}}_{a_{1}\cdots a_{j-1}ea_{j+1}\cdots a_{p}}(\phi_{1})~
\stackrel{2}{\partial_{a_{j}}}h^{e}(\phi_{2})\cr
\ + & \ \bigl[{\cal G}(\lambda\omega\stackrel{2}{\partial}-{i\over 2}\stackrel
{1}{\partial}\omega\stackrel{1}{\partial})-{\cal G}(\lambda\omega\stackrel
{2}{\partial}+\stackrel{1}{\partial}\omega\stackrel{2}{\partial})\bigr]\cdot\cr
\ \cdot & \ T^{b_{1}\cdots b_{q}}_{a_{1}\cdots a_{p}}(\phi_{1})~
\stackrel{2}{\partial_{f}}h^{e}(\phi_{2})~{\bar c}_{e}c^{f}\biggr)
\vert_{\phi_{1,2}=\phi}\cr
\ \ast_{e} & \ {\bar c}_{b_{1}}\cdots{\bar c}_{b_{q}}~c^{a_{1}}\cdots
c^{a_{p}}\cr}\eqn\lasagna$$
with ~$\lambda\omega\partial\equiv\lambda_{a}\omega^{ab}\partial_{b}$,~etc., and
~${\cal F}$~and ~${\cal G}$~defined in ~\impaziente~and~\dizionario,
respectively. Some details of the calculations are given in appendix
C. The point to be noticed is that, because of the term involving
~${\bar c}_{e}c^{f}$, the bracket with ~$\Haq$~maps a (p,q)-tensor
to the sum of a (p,q) and a (p+1,q+1)-tensor. Upon projection on the
~$\lambda=0$~surface the (p+1,q+1)-piece vanishes. In fact, putting
~$\lambda=0$, the two terms inside the square brackets of the last 
term on the RHS of~\lasagna~cancel because ~${\cal G}$~is an even function.
(This would not be the case for the Hamiltonians~$\Haq_{B\pm}$.)~
Since the tensor structure of ~${\cal P}\{\Haq,{\widehat T}\}_{emb}$~matches
that of ~${\widehat T}$, we may strip off the ~${\bar c}$'s and the
$c$'s. This leaves us with 
$$\eqalign{{L}_{h}T^{b_{1}\cdots b_{q}}_{a_{1}\cdots a_{p}}\ (&\ \phi)= \bigl\{T^{b_{1}\cdots b_{q}}_{a_{1}\cdots a_{p}}(\phi),H\bigr\}
_{mb}-\cr
\ - & \ \sum_{j=1}^{q}{\sin[{\hbar\over 2}\stackrel{1}{\partial}\omega
\stackrel{2}{\partial}]\over [{\hbar\over 2}\stackrel{1}{\partial}\omega
\stackrel{2}{\partial}]}T^{b_{1}\cdots b_{j-1}e b_{j+1}\cdots b_{q}}
_{a_{1}\cdots a_{p}}(\phi_{1})\stackrel{2}{\partial_{e}}h^{b_{j}}(\phi_{2})\vert
_{\phi_{1,2}=\phi}+\cr 
\ + & \ \sum_{j=1}^{p}{\sin[{\hbar\over 2}\stackrel{1}{\partial}\omega
\stackrel{2}{\partial}]\over [{\hbar\over 2}\stackrel{1}{\partial}\omega
\stackrel{2}{\partial}]}T^{b_{1}\cdots b_{q}}
_{a_{1}\cdots a_{j-1}e a_{j+1}\cdots a_{p}}(\phi_{1})\stackrel{2}
{\partial_{a_{j}}}h^{e}(\phi_{2})\vert
_{\phi_{1,2}=\phi}\cr}\eqn\casale$$ 
The index structure in ~\casale~ is the same as in the classical case
but the tensor components are multiplied by ~$\partial_{a}h^{b}$~by
means of a new nonlocal product. In particular for a 1-form~$\varrho_{a}$
~one finds
$${L}_{h}\varrho_{a}=\bigl\{\varrho_{a},H\bigr\}_{mb}+
{\sin[{\hbar\over 2}\stackrel{1}{\partial}\omega\stackrel{2}{\partial}]
\over[{\hbar\over 2}\stackrel{1}{\partial}\omega\stackrel{2}{\partial}]}
\varrho_{b}(\phi_{1})
\stackrel{2}{\partial_{a}}h^{b}
(\phi_{2})\vert_{\phi_{1,2}=\phi}\eqn\vicomoscano$$
Taking the partial derivative on both sides of eq.~\tullio~ and comparing
to ~\vicomoscano~shows that the derivative of zero-forms,
~$\varrho_{a}\equiv \partial_{a}\varrho$, evolves as a 1-form, exactly 
as it happened in classical mechanics. That would 
not be the case for any other bosonic super-Hamiltonian different from the
symmetric one ~$\Haq_{B0}$. This is a further reason to choose
this form for the quantum hamiltonian vector field.
\par
Finally we have to ask in which sense the quantum Lie derivatives form
a closed algebra. In our approach ~the action of~$L_{h}$~
on  tensor-fields is represented 
as the Moyal bracket of a certain super-Hamiltonian with those
tensors; therefore the algebra of the ~$L_{h}$'s~closes by construction
on the space of {\it all} generating functions on ~${\cal M}_{8N}$.
However, restricting the generating functions to the~$\Haq$-type, there is no
guarantee that the algebra still closes. In eq.~\ascensore~ we have seen that
the ~$emb$-algebra of the bosonic parts ~$\Haq_{B}$~closes nevertheless,
and it is also known\refmark{11} that for ~$\hbar=0$~the ~$epb$-bracket
algebra of the full ~${\widetilde{\cal H}}={\widetilde{\cal H}}_{B}+
{\widetilde{\cal H}}_{F}$~is closed as well. It turns out that for
~$\hbar\ne 0$~the algebra does not close on the space of the super-Hamiltonians
~$\Haq$, but only on a slightly larger one. To see this, let us try to
generalize ~\ascensore~by adding the fermionic piece
~$\Haq_{F}$~and let us look at the term~$\bigl\{\Haq_{F}[G_{1}],\Haq_{F}[G_{2}]
\bigr\}_{emb}$. It contains a term with four ghosts
$$-{\bar c}_{a}{\bar c}_{d}c^{b}c^{c}\bigl\{\partial_{b}h^{a}_{1,\hbar},
\partial_{c}h^{d}_{2,\hbar}\bigr\}_{emb}\eqn\settetrentadue$$
which prevents the algebra from closing on another~$\Haq$. The 
term~\settetrentadue~vanishes in the classical limit and also for ~$\lambda_{a}
=0$. But, of course, in order to make sure that the transformations
~$L_{h}{\widehat T}=\bigl\{{\widehat T},\Haq\bigr\}_{emb}$~close,
we are not allowed to set ~$\lambda=0$,~${\bar c}=0$~before having acted
on ~${\widehat T}$. Let us denote by ~${\cal C}$~the subspace of
~$Fun({\cal M}_{8N})$~consisting of the functions~${\widetilde\Gamma}$~which
can be obtained by taking repeated ~$em$-brackets of the super-Hamiltonians
~$\Haq$. Then the algebra of the (generalized) Lie derivatives
~$L({\widetilde\Gamma})=\bigl\{\cdot, {\widetilde\Gamma}\bigr\}_{emb}$~closes trivially on the enlarged
space~${\cal C}$, which contains the old super-Hamiltonians as well as new
types of functions with more than one~${\bar c}c$-pair. It seems quite
natural to call the transformation~$L({\widetilde\Gamma})=\bigl\{\cdot, {\widetilde\Gamma}
\bigr\}_{emb}$,
for ~{\it any} ~${\widetilde\Gamma}\in{\cal C}$, a quantum
canonical transformation, even though there are many more~
${\widetilde\Gamma}$'s~
than classical generating functions ~$G\in Fun({\cal M}_{2N})$. In fact,
every ~${\widetilde\Gamma}\in{\cal C}$~is of the form
$${\widetilde\Gamma}=\Haq[G]+\Delta{\widetilde\Gamma}~~~,~~~\Delta{\widetilde
\Gamma}=O(\hbar)\eqn\settetrentatre$$
\ie,~for ~$\hbar\rightarrow 0$~each~~${\widetilde\Gamma}$~equals a conventional
classical super-Hamiltonian for some ~$G\in Fun({\cal M}_{2N})$, but there
can be many ~${\widetilde\Gamma}$'s which have the same classical limit.
In the classical case a (symplectic) diffeomorphism is equivalent to a vector
field, and two consecutive transformations are again equivalent
to a vector field, the Lie bracket of the original ones. In the deformed
case the product of two canonical transformations (in the "narrow" sense
of~$\Haq$) can be something more general than a vector field. In fact,
applying a term like~\settetrentadue~to a tensor ~${\widehat T}$~leads
(after projection) to tensors of the type
$$T^{b_{1}\cdots b_{q}}_{cda_{3}\cdots a_{p}}(\phi)~\ast_{e}\bigl\{\partial
_{a_{1}}h^{c}_{1,\hbar},\partial_{a_{2}}h^{d}_{2,\hbar}\bigr\}_{emb}\vert
_{\lambda=0}\eqn\settetrentaquattro$$
This is a contribution to ~$L({\widetilde\Gamma}){\widehat T}$~which cannot
be parametrized by a single vector field. It should be compared with the RHS
of ~\casale. The novel feature is that we can act with 
~$\partial_{a}h^{b}_{\hbar}$~on more than one tensor index. In general
the ~$\Delta{\widetilde\Gamma}$-terms induce rather complicated terms at the
level of tensor components and we shall give no explicit
formulas here. It has to be remarked, however, that the functions ~${\widetilde
\Gamma}$,~though more complicated than~$\Haq$, still have a very particular
form, since they are invariant under the full ~$ISp(2)$~group:~$\bigl\{
{\widetilde\Gamma},\Omega\bigr\}_{emb}=0$. This follows from the fact that the
~${\widetilde\Gamma}$'s are obtained as ~$em$-brackets of~$\Haq$'s.
\par
Classically tensors are representations of the
group of diffeomorphisms on ~${\cal M}_{2N}$~which
is essentially the same as  ~$Fun({\cal M}_{2N})$~in the symplectic case. In our approach 
"{\it quantum p-forms}" are representations of the algebra generated by
~$\bigl\{{\widetilde\Gamma},\cdot\bigr\}$, which is larger than\break $Fun({\cal M}
_{2N})$. Therefore specifying a classical canonical transformation does not
uniquely fix a quantum canonical transformation for ~$p\geq 1$~. We encounter
a kind of {\it "holonomy effect}" on ~${\cal C}$. What we mean is that
we can associate to ~$G\in Fun({\cal M}_{2N})$~directly a ~$\Haq[G]\equiv
{\widetilde\Gamma}$~or we can reach a quantum transformation associated 
to the same classical~$G$~via some intermediate transformations ~
$G_{1}$~and~$G_{2}$, and 
the ~${\widetilde\Gamma}$~we shall
obtain will be different from the previous one. This {\it "holonomy effect"}
might be at the heart of  QM irrespective of the quantum-tensor
calculus we choose. It may be that this  over-all attempt to understand 
the geometry of QM, by defining quantum-forms and similar structures, 
sheds some light on the old problem of the {\it non-local} nature of quantum
mechanics. Work is in progress 
on this issue\Ref\MSA{E.Gozzi, M.Reuter, work in progress} especially
in the direction of getting a ~$\Haq_{F}$~ from more physical requirements. 
\par
Table 2  summarizes  the
~$emb$-representations of the various {\it quantum}-tensor manipulations 
and it should be compared with table 1 where the {\it classical}-tensor
manipulations were summarized. In view of the above it should be kept 
in mind that there are more general quantum Lie derivatives than the
~$L_{h}$~displayed in the table.
\chapter{CONCLUSIONS}
In this paper we have made a proposal for a quantum exterior calculus
which stays as close as possible to the classical one. We have been able to find
a calculus in which the pointwise product of functions was deformed, but 
the wedge-product was left unaltered. The quantum deformed hamiltonian vector
field and Lie derivative are , in some respect, surprisingly similar to their classical counterparts, the only difference being the 
non-locality creeping in through non-local star products and "lattice"
quantities. Moreover some extra variables,~$\lambda,c,{\bar c}$,~appeared 
which were needed in order to unfold the geometrical properties of
quantum mechanics. A crucial role is played by the auxiliary variable
~$\lambda_{a}$. Though formally equivalent to the response 
field\refmark{20} of statistical mechanics, in quantum mechanics its physics seems to be much more 
involved. In particular it seems to be at the heart of the "foamy" structure of
quantum phase-space: it partitions phase-space into Planck cells of size
~$\Delta p~\Delta q\sim\hbar$.
\par
Leaving aside the interesting mathematical properties of the new variables
~$\lambda_{a}$,~$c^{a}$,~and ~${\bar c}_{a}$, it is tempting to speculate 
about their possible role in physics. It is intriguing that the theory might
have a consistent interpretation even away from the ~$\lambda=0$-surface.
One could envisage a situation in which, allowing for ~$\lambda\ne 0$,
smoothens out certain (space-time) singularities present for ~$\lambda=0$.
It seems very likely that this would allow for an evolution of pure states
into mixed states as it might possibly occur in the late stage of black
hole evaporation\Ref\hawk{S.Hawking, Com.Math.Phys. 43 (1975) 199;\nextline
J.Ellis et al. Nucl.Phys.B241 (1984) 381}. It is also conceiveable that the inclusion of the
fermionic sector improves the renormalizability properties of some models.
\par
An important step towards a physical understanding of the extended theory 
presented here would be a reformulation of the formalism in terms of 
wave-functions and Hilbert spaces. To achieve that it is probably mandatory 
to choose the fermionic sector\break (\ie,~$\Haq_{F}$)~in such a way that the 
holomorphic decomposition is manifest also at the level of ~$c$~ and ~
${\bar c}$.
In our present formulation this is not yet the case. A related problem 
is that the universal supersymmetry (SUSY) present in the classical case
\refmark{11,19}, contrary to the ~$ISp(2)$, does not seem to
survive the quantum deformation. We have shown in ref.[19] how that classical SUSY was strictly
related to the concept of classical ergodicity\Ref\arno{V.I.Arnold and A.Avez,
"{\it Ergodic problems of classical mechanics"},\nextline
Benjamin, New York, 1968} and how it could nicely reproduce the classical 
KMS conditions\Ref\gall{G.Gallavotti and E.Verboven, Nuovo Cimento 28
(1975) 274 }. It would be interesting to find
(possibly for a different ~$\Haq_{F}$) a {\it deformed}
SUSY which could be used to study quantum ergodicity\Ref\von{J.von Neumann,
Z.Phys. 57 (1929) 30} and the quantum KMS conditions.
\par
In conclusion we can say that in our approach a tensor calculus is selected by
choosing a specific quantum-Lie-derivative (or super-Hamiltonian).
Its bosonic part, ~${\cal H}_{B}$, is essentially unique and represents Moyal's
deformed hamiltonian vector field. For the
fermionic part ~${\Haq}_{F}$~many choices are possible a priori and future
work will have to show their physical relevance, and their relation to other
approaches.\Ref\wes{J.Wess, B.Zumino, O.Ogievetskii, W.B.Schmidke, Munich
preprint 93-192;\nextline
J.Wess et al., Munich preprint 93-0194}
\ack
This research has been supported in part by grants from INFN, MURST and NATO.
M.R. acknowledges the hospitality of the Dipartimento di Fisica Teorica,
Universit\`a di Trieste while this work was in progress.
\Appendix{A}
In this appendix we solve eq.~\vho~and show that the original equation
~\sangiovanni~has no solution.
The RHS of eq.~\sangiovanni~reads, using ~\vittus,~
$$\eqalign{\bigl\{H(\phi),\varrho(\phi)\bigr\}_{mb} \ = & \ {2\over\hbar}
\sin\bigl[{\hbar\over 2}\omega^{ab}\stackrel{1}{\partial_{a}}\stackrel{2}
{\partial_{b}}\bigr]H(\phi_{1})~\varrho(\phi_{2})\vert_{\phi_{1,2}=\phi}\cr
\ = & \ {2\over\hbar}\sum_{m=0}^{\infty}{(-1)^{m}\over (m+1)!}~\bigl({\hbar\over
2}\bigr)^{2m+1}\omega^{a_{1}b_{1}}\cdots\omega^{a_{2m+1}b_{2m+1}}\partial
_{a_{1}}\cdots\partial_{a_{2m+1}}H(\phi)\cr
\ ~~~ & \ \cdot \partial_{b_{1}} \cdots
\partial_{b_{2m+1}}\varrho(\phi)\cr}\eqn\vita$$
In a similarly way  the LHS of eq.~\sangiovanni~ follows from ~\verona~
and ~\pordenone:
$$\eqalign{\bigl\{\Haq_{B}(\lambda,\phi) \ , & \ \varrho(\phi)\bigr\}_{emb}  =
2\sin\bigl[{1\over 2}\bigl({\partial\over\partial\phi^{a}_{1}}{\partial\over
\partial\lambda_{2a}}-{\partial\over\partial\lambda_{1a}}{\partial\over\partial
\phi_{2}^{a}}\bigr)\bigr]\Haq_{B}(\lambda_{1},\phi_{1})\varrho(\phi_{2})\vert_
{\phi_{1,2}=\phi\atop \lambda_{1,2}=\lambda}\cr
\ = & \ -2\sin\bigl[{1\over 2}{\partial\over\partial\lambda_{a}}{\partial\over
\partial{\tilde\phi}^{a}}\bigr]{\Haq_{B}}(\lambda,\phi)~\varrho({\tilde\phi})
\vert_{{\tilde\phi}=\phi}\cr
\ = & \ -2\sum_{m=0}^{\infty}{(-1)^{m}\over (2m+1)!}\bigl({1\over 2}\bigr)
^{2m+1}{\partial\over\partial\lambda_{b_{1}}}\cdots{\partial\over\partial
\lambda_{b_{2m+1}}}{\widetilde{\cal H}}_{B}(\lambda,\phi)
~\partial_{b_{1}}\cdots\partial_{b_{2m+1}}\varrho(\phi)\cr}\eqn\morte$$
As ~$\varrho(\phi)$~is an arbitrary function, we can compare the coefficients
of ~$\partial_{b_{1}}\cdots \partial_{b_{2m+1}}\varrho$~in ~\morte~and
~\vita, so that eq.~\sangiovanni~is equivalent to the set of equations
$$\eqalign{\ \bigl( & \ \hbar\omega^{a_{1}b_{1}}\partial_{b_{1}}\bigr)
\bigl(\hbar\omega^{a_{2}b_{2}}\partial_{b_{2}}\bigr)\cdots
\bigl(\hbar\omega^{a_{2m+1}b_{2m+1}}\partial_{b_{2m+1}}\bigr)H(\phi)=\cr
\ = & \ \hbar {\partial\over\partial\lambda_{a_{1}}}
{\partial\over\partial\lambda_{a_{2}}}\cdots{\partial
\over\partial\lambda_{a_{2m+1}}}~\Haq_{B}(\lambda,\phi)\cr}\eqn\genio$$
where ~$m=0,1,2,\cdots$. Before studying ~\genio~in general,
let us check its classical limit. For ~$\hbar\rightarrow 0$~one finds
$${\partial\over\partial\lambda_{a}}{\widetilde{\cal H}}_{B}=\omega^{ab}
\partial_{b}H=h^{a}\eqn\maniaco$$
$${\partial\over\partial_{a_{1}}}\cdots{\partial\over\partial_{a_{2m+1}}}
{\widetilde{\cal H}}_{B}=0~~,~~m=1,2,3,\cdots\eqn\peperoni$$
The unique solution to ~\maniaco,~\peperoni~is
$${\widetilde{\cal H}}_{B}=\lambda_{a}h^{a}\eqn\pizza$$
which, as expected, coincides with ~\paris.
\par
Returning to ~$\hbar\ne 0$, it is quite obvious that the infinite system
of equations in ~\genio~has no solution. Setting ~$m=0$~we find ~\maniaco~
again. Its solution~\pizza~has vanishing, second, third, etc. derivatives
with respect to ~$\lambda$. Therefore it does not solve ~\genio~ for ~$m=1,2,
\cdots$~and for a generic Hamiltonian ~$H$. Hence, as discussed in section 5, we
only require the weaker condition~\vho~ which is equivalent to ~\genio~ with ~
$\lambda$~put to zero on the RHS of ~\genio~after the derivatives have been
taken.
\par
We assume that~$\Haq(\lambda,\phi)$~is analytic in ~$\lambda$~and make
the ansatz
$$\Haq_{B}(\lambda,\phi)=\sum_{l=0}^{\infty}{1\over l!}~\Theta_{(l)}^{a_{1}\cdots
a_{l}}(\phi)~\lambda_{a_{1}}\cdots\lambda_{a_{l}}\eqn\radio$$
with the symmetric coefficients
$$\Theta_{(l)}^{a_{1}\cdots a_{l}}(\phi)=
{\partial\over\partial\lambda_{a_{1}}}\cdots{\partial\over\partial
\lambda_{a_{l}}}\Haq_{B}(\lambda,\phi)\vert_{\lambda=0}\eqn\televisione$$
Inserting~\radio~into~\genio~the coefficients ~$\Theta_{(l)}$,~for
even values of ~$l$, are left unconstrained and those with odd values
~$l\equiv 2m+1$,~$m=0,1,2,\cdots$~are fixed to be
$$\Theta_{(2m+1)}^{a_{1}\cdots a_{2m+1}}(\phi)={1\over\hbar}\bigl(\hbar
\omega^{a_{1}b_{1}}\partial_{b_{1}}\bigr)\cdots\bigl(\hbar\omega^{a_{2m+1}
b_{2m+1}}\partial_{b_{2m+1}}\bigr)H(\phi)\eqn\giradischi$$
For the choice~$\Theta_{(l)}=0$~for ~$l$~even, ~\radio~with~\giradischi~
yields
$$\eqalign{\Haq_{B}(\lambda,\phi) \ = & \ {1\over\hbar}\sum_{m=0}^{\infty}
{1\over (2m+1)!}\bigl(\hbar\lambda_{a}\omega^{ab}\partial_{b}\bigr)^{2m+1}
H(\phi)\cr
\ = & \ {1\over\hbar}\sinh\bigl[\hbar\lambda_{a}\omega^{ab}\partial_{b}\bigr]
H(\phi)\cr}\eqn\registratore$$
As the~$\Theta_{(l)}$'s with ~$l$~ even are not fixed by ~\genio~we are free
to add to ~\registratore~any function which is even in ~$\lambda$. This 
leads to eq.~\castellucchio~given in section 5.
\endpage
\Appendix{B}
In this appendix we show that the conditions~{\bf F1}~and~{\bf F2}~of section
6 imply the super-Hamiltonian of eq.~\ruth. We start from the ansatz
$$\Haq={\cal F}(\lambda\omega\partial)H(\phi)+i{\bar c}_{a}W^{a}_{b}
(\lambda,\phi)~c^{b}\eqn\biuno$$
and we require ~$\Haq$~to be BRS invariant. Applying the
transformations~\bologna~to ~\biuno~one finds
$$\delta\Haq=\epsilon c^{b}{\cal F}(\lambda\omega\partial)~\partial_{b}H(\phi)
-\epsilon\lambda_{a}W^{a}_{b}(\lambda,\phi)c^{b}-i\epsilon{\bar c}_{a}
\partial_{c}W^{a}_{b}(\lambda,\phi)~c^{c}~c^{b}\eqn\bidue$$
From ~$\delta\Haq=0$~it follows that
$$\partial_{c}W^{a}_{b}-\partial_{b}W^{a}_{c}=0\eqn\bitre$$
and
$${\cal F}(\lambda\omega\partial)~\partial_{b}H(\phi)=\lambda_{a}W^{a}_{b}
(\lambda,\phi)\eqn\biquattro$$
Assuming that ~${\cal M}_{2N}$~is topologically trivial, eq.~\bitre~
implies that
$$W^{a}_{b}(\lambda,\phi)=\partial_{b}W^{a}(\lambda,\phi)=\omega^{ac}
\partial_{b}U_{c}(\lambda,\phi)\eqn\bisei$$
for some function ~$W^{a}\equiv\omega^{ac}U_{c}$. Inserting~\bisei~into
~\biquattro~yields
$$\bigl(\lambda_{a}\omega^{ac}\partial_{c}\bigr)\partial_{b}{{\cal F}
(\lambda\omega\partial)\over (\lambda\omega\partial)}H(\phi)=\lambda_{a}
\omega^{ac}\partial_{b}U_{c}(\lambda,\phi)\eqn\bisette$$
On the LHS of this equation ~$\lambda$~appears only in the combination ~
$\lambda_{a}\omega^{ac}\partial_{c}$, so the same must be true for the RHS also.
This implies that ~$U_{c}$~is a gradient: ~$U_{c}(\lambda,\phi)=\partial_{c}
U(\phi,\lambda)$. Inserting this into ~\bisette~ we find that (up to an
irrelevant constant)
$$U(\lambda,\phi)={{\cal F}(\lambda\omega\partial)\over (\lambda\omega\partial)}
H(\phi)\eqn\biotto$$
and therefore
$$W^{a}_{b}(\lambda,\phi)=\omega^{ac}~\partial_{b}\partial_{c}~{{\cal F}(\lambda
\omega\partial)\over (\lambda\omega\partial)}H(\phi)\eqn\binove$$
Eq.~\biuno~with ~\binove~is the result~\ruth~given in section 6. Note that
~$W^{a}_{a}(\lambda,\phi)=0$~so that, as in the classical case, ~
${\widehat{\bar c}}_{a}~W^{a}_{b}~{\widehat c}^{b}$~is free from ordering 
ambiguities.
\Appendix{C}
In this appendix we give some details of the derivation of eq.~\lasagna.
The evaluation of ~$\bigl\{\Haq,{\widehat T}\bigr\}_{emb}$, with
~${\widehat T}$~defined in ~\lianas,~proceeds by repeated application
of eq.~\mantova. In a first step we write
$$\bigl\{\Haq,{\widehat T}\bigr\}_{emb}={\cal R}_{1}+{\cal R}_{2}+
{\cal R}_{3}\eqn\ciuno$$
with
$$\eqalign{{\cal R}_{1} \ = & \ \bigl\{\Haq_{B},T^{b_{1}\cdots b_{q}}_{a_{1}
\cdots a_{p}}\bigr\}_{emb}\ast_{e}{\bar c}_{b_{1}}\cdots {\bar c}_{b_{q}}~
c^{a_{1}}\cdots c^{a_{p}}\cr
{\cal R}_{2} \ = & \ T^{b_{1}\cdots b_{q}}_{a_{1}\cdots a_{p}}\ast_{e}~
\biggl[{\bar c}_{b_{1}}\cdots {\bar c}_{b_{q}}\ast_{e}\bigl\{\Haq_{F},c^{a_{1}}
\cdots c^{a_{p}}\bigr\}_{emb}+\bigl\{\Haq_{F},{\bar c}_{b_{1}}\cdots 
{\bar c}_{b_{q}}\bigr\}_{emb}\ast_{e}c^{a_{1}}\cdots c^{a_{p}}\biggr]\cr
{\cal R}_{3} \ = & \ \bigl\{\Haq_{F},T^{b_{1}\cdots b_{q}}_{a_{1}\cdots a_{p}}
\bigr\}_{emb}\ast_{e}{\bar c}_{b_{1}}\cdots {\bar c}_{b_{q}}c^{a_{1}}
\cdots c^{a_{p}}\cr}\eqn\cidue$$
Using ~\battipaglia~and ~\viadana~we obtain for the first piece
$${\cal R}_{1}=i\biggl[{\cal F}\bigl([\lambda+{i\over 2}\stackrel{1}{\partial}
]\omega\stackrel{2}{\partial}\bigr)-{\cal F}\bigl([\lambda-{i\over 2}\stackrel
{1}{\partial}]\omega\stackrel{2}{\partial}\bigr)\biggr]~T^{b_{1}\cdots b_{q}}
_{a_{1}\cdots a_{p}}(\phi_{1})H(\phi_{2})\vert_{\phi_{1,2}=\phi}~
{\bar c}_{b_{1}}\cdots {\bar c}_{b_{q}}c^{a_{1}}\cdots c^{a_{p}}\eqn\citre$$
which is the first term on the RHS of~\lasagna.
For ~${\cal R}_{2}$~we need:
$$\eqalign{\bigl\{\Haq_{F},c^{a_{1}}\cdots c^{a_{p}}\bigr\}_{emb}
\ = & \ \sum_{j=1}^{p}~c^{a_{1}}\cdots c^{a_{j-1}}\ast_{e}\bigl\{
\Haq_{F},c^{a_{j}}\bigl\}\ast_{e}c^{a_{j+1}}\cdots c^{a_{p}}\cr
\ = & \ -\sum_{j=1}^{p}~\partial_{b}h^{a_{j}}_{\hbar}~c^{a_{1}}\cdots
c^{a_{j-1}}c^{b}c^{a_{j+1}}\cdots c^{a_{p}}\cr}\eqn\ciquattro$$
and a similar formula for ~${\bar c}_{b_{1}}\cdots {\bar c}_{b_{q}}$.
This leads to
$$\eqalign{{\cal R}_{2}= \ \bigl [ & \ \sum_{j=1}^{p}T^{b_{1}\cdots b_{j-1}e
b_{j+1}\cdots b_{q}}_{a_{1}\cdots a_{p}}\ast_{e}\partial_{e}h^{b_{j}}_{\hbar}\cr
\ - & \ \sum_{j=1}^{q}~T^{b_{1}\cdots b_{q}}_{a_{1}\cdots a_{j-1}e
a_{j+1}\cdots a_{p}}\ast_{e}\partial_{a_{j}}h_{\hbar}^{e}\bigr]{\bar c}_{b_{1}}
\cdots {\bar c}_{b_{q}}c^{a_{1}}\cdots c^{a_{p}}\cr}\eqn\cicinque$$
Because the quantum lift ~$h_{\hbar}^{a}={\cal G}(\lambda\omega\partial)h^{a}$~
depends on ~$\lambda$, the star products between the tensor components and
~$\partial_{a}h^{b}_{\hbar}$~are non-trivial. We use~\viadana~again:
$$T^{\cdots}_{\cdots}(\phi)\ast_{e}
\partial_{a}h^{b}_{\hbar}(\lambda,\phi)={\cal G}(\lambda\omega\stackrel
{2}{\partial}+{i\over 2}\stackrel{1}{\partial}\omega\stackrel{2}{\partial}
)~T^{\cdots}_{\cdots}(\phi_{1})~\partial_{a}h^{b}(\phi_{2})\vert_{\phi_{1,2}
=\phi}\eqn\cisei$$
Eqs.~\cicinque~with ~\cisei~yields the second and the third term on the
RHS of eq.~\lasagna.
The term ~${\cal R}_{3}$~ is a typical non-classical feature. The
bracket~$\bigl\{\Haq_{F},T^{\cdots}_{\cdots}\bigr\}_{emb}$~is non-zero 
only because ~$h^{a}_{\hbar}$~depends on ~$\lambda$. In the classical limit 
this bracket vanishes. We have
$${\cal R}_{3}=i{\bar c}_{e}c^{f}\bigl\{{\cal G}(\lambda\omega\partial)
\partial_{f}h^{e}(\phi),T^{b_{1}\cdots b_{q}}_{a_{1}\cdots a_{p}}(\phi)
\bigr\}_{emb}\ast_{e}{\bar c}_{b_{1}}\cdots {\bar c}_{b_{q}}c^{a_{1}}
\cdots c^{a_{p}}\eqn\cinove$$
Applying eq.~\viadana~this yields the last term on the RHS of eq.~\lasagna.
\refout
\eject

%






\def\bigstrut#1#2{\hbox{\vrule height#1pt depth#2pt width0pt}} 







\setbox0=\vbox{
	\halign{\hfil#\hfil&#\cr
		Lie-deriv.&\cr
		on T=$F^{(p)}$,$V^{(p)}$&\cr
}}
\relax
\noindent\hfill
\vbox{\tabskip=0pt \offinterlineskip 
	\def\tablerule{\noalign{\hrule}}
\halign to 450pt {
	\strut#& 				
	\vrule#\tabskip=1em plus 2em&		
	\hfil#\hfil& 				
	\vrule#& 				
	\hfil#\hfil& 				
	\vrule#& 				
	\hfil#\hfil& 				
	\vrule#\tabskip=0pt 					\cr
	\tablerule
	&&
	\multispan5 	
	\hidewidth
	\bigstrut{16}{12}		Table 1 
	\hidewidth
	&							\cr
	\tablerule
	\bigstrut{14}{10}&&
	$\bigcirc$&&
	\bf Classical Cartan's Rules&&
	\bf $\{\cdot,\cdot\}_{epb}$-Rules&		\cr	
	\tablerule
	\bigstrut{20}{20}&&
	vector-fields&&
	$v=v^{a}\partial_{a}$&&
	$ v=v^{a}{\bar c}_{a}$&			\cr	
	\tablerule
	\bigstrut{20}{20}&&
	1-forms&&
	$\alpha=\alpha_{a}d\phi^{a}$&&
	$\alpha=\alpha_{a}c^{a}$&			\cr	
	\tablerule
	\bigstrut{20}{20}&&
	p-forms&&
	$F^{(p)}={1\over p!}F_{a_{1}\cdots a_{p}}d\phi^{a_{1}}\wedge
\cdots\wedge d\phi^{a_{p}}$&&
	$ F^{(p)}={1\over p!} F_{a_{1}\cdots a_{p}}c^{a_{1}}\cdots
c^{a_{p}}$&							\cr	
	\tablerule
	\bigstrut{20}{20}&&
	p-vectors&&
	$V^{(p)}={1\over p!}V^{a_{1}\cdots a_{p}}\partial_{a_{1}}\wedge\cdots
\wedge\partial_{a_{p}}$&&
	$V^{(p)}={1\over p!}V^{a_{1}\cdots a_{p}}{\bar c}_{a_{1}}\cdots
{\bar c}_{a_{p}}$&							\cr	
	\tablerule
	\bigstrut{20}{20}&&
	ext.~deriv.&&
	$dF^{(p)}$&&
	$i\bigl\{Q, F^{(p)}\bigr\}_{epb}$&	      \cr	
	\tablerule
	\bigstrut{20}{20}&&
	int.~product&&
	$i(v)F^{(p)}$&&
	$i\bigl\{ v, F^{(p)}\bigr\}_{epb}$&		\cr	
	\tablerule
	\bigstrut{20}{20}&&
	Ham.vec.field&&
	$h=\omega^{ab}\partial_{b}H\partial_{a}$&&
	$h=i\omega^{ab}\partial_{b}H\lambda_{a}$&			\cr	
	\tablerule
	\bigstrut{20}{20}&&
	Lie-deriv.&&
	$l_{h}=di_{h}+i_{h}d$&&
	$i{\widetilde{\cal H}}=i\lambda_{a}h^{a}-{\bar c}_{a}
\partial_{b}h^{a}c^{b}$&            \cr
        \tablerule
	\bigstrut{27}{27}&&
	\lower9pt\box0&&
	$l_{h}T$&&
	$-\{{\widetilde{\cal H}}, T\}_{epb}$&			\cr	
	\noalign{\hrule\smallskip}
}}
\hfill\hfill
\endpage


%






\def\bigstrut#1#2{\hbox{\vrule height#1pt depth#2pt width0pt}} 







\setbox0=\vbox{
	\halign{\hfil#\hfil&#\cr
		Lie-deriv.&\cr
		on T=$F^{(p)}$,$V^{(p)}$&\cr
}}

\relax
\noindent\hfill
\vbox{\tabskip=0pt \offinterlineskip 
	\def\tablerule{\noalign{\hrule}}
\halign to 460pt {
	\strut#& 				
	\vrule#\tabskip=1em plus 2em&		
	\hfil#\hfil& 				
	\vrule#& 				
	\hfil#\hfil& 				
	\vrule#& 				
	\hfil#\hfil& 				
	\vrule#\tabskip=0pt 					\cr
	\tablerule
	&&
	\multispan5 	
	\hidewidth
	\bigstrut{16}{12}		Table 2
	\hidewidth
	&							\cr
	\tablerule
	\bigstrut{14}{10}&&
	$\bigcirc$&&
	\bf Cartan's Rules&&
	{\it quantum}-{\bf Cartan's Rules}&		\cr	
	\tablerule
	\bigstrut{20}{20}&&
	vector-fields&&
	$v=v^{a}\partial_{a}$&&
	$ {\widehat v}=v^{a}{\bar c}_{a}$&				\cr	
	\tablerule
	\bigstrut{20}{20}&&
	1-forms&&
	$\alpha=\alpha_{a}d\phi^{a}$&&
	${\widehat\alpha}=\alpha_{a}c^{a}$&      \cr	
	\tablerule
	\bigstrut{20}{20}&&
	forms&&
	$F^{(p)}={1\over p!}F_{a_{1}\cdots a_{p}}d\phi^{a_{1}}\wedge
\cdots\wedge d\phi^{a_{p}}$&&
	$ {\widehat F}^{(p)}={1\over p!} F_{a_{1}\cdots a_{p}}c^{a_{1}}\cdots
c^{a_{p}}$&							\cr	
	\tablerule
	\bigstrut{20}{20}&&
	tensors&&
	$V^{(p)}={1\over p!}V^{a_{1}\cdots a_{p}}\partial_{a_{1}}\wedge\cdots
\wedge\partial_{a_{p}}$&&
	${\widehat V}^{(p)}={1\over p!}V^{a_{1}\cdots a_{p}}{\bar c}_{a_{1}}\cdots
{\bar c}_{a_{p}}$&							\cr	
	\tablerule
	\bigstrut{20}{20}&&
	ext-deriv.&&
	$dF^{(p)}$&&
	$i\bigl\{ Q, {\widehat F}^{(p)}\bigr\}_{emb}$&	\cr	
	\tablerule
	\bigstrut{20}{20}&&
	int.product&&
	$i(v)F^{(p)}$&&
	$i\bigl\{ v, {\widehat F}^{(p)}\bigr\}_{emb}$&		\cr	
	\tablerule
	\bigstrut{20}{20}&&
	Ham-Vec-field&&
	$h=\omega^{ab}\partial_{b}H\partial_{a}$&&
	$ h^{a}_{\hbar}\equiv{\sinh(\hbar\lambda\omega\partial)\over (\hbar
\lambda\omega\partial)}h^{a}(\phi)$&			\cr	
	\tablerule
	\bigstrut{20}{20}&&
	Lie-deriv.&&
	$L_{h}=di_{h}+i_{h}d$&&
	$i{\widetilde{\cal H}^{\hbar}}=i\lambda_{a} h^{a}_{\hbar}-
{\bar c}_{a}\partial_{b} h^{a}_{\hbar}c^{b}$&            \cr
        \tablerule
	\bigstrut{27}{27}&&
	\lower9pt\box0&&
	$L_{h}T$&&
	$-\{{\widetilde{\cal H}^{\hbar}}, T\}_{emb}$&	\cr	
	\noalign{\hrule\smallskip}
}}
\hfill\hfill
\endpage
\vfil\eject
\bye